# Perpendicular Magnetic Anisotropy in Conducting $NiCo_2O_4$ Films from Spin-Lattice Coupling

Corbyn Mellinger[1], Jace Waybright[1,2], Xiaozhe Zhang[1,3], Caleb Schmidt[1], and Xiaoshan Xu[1,4]

[1] Department of Physics and Astronomy, University of Nebraska, Lincoln, Nebraska 68588, USA

[2] Department of Physics, South Dakota State University, Brookings, SD 57007 USA

[3] School of Materials Science and Engineering, Xi'an Polytechnic University, Xi'an 710049, Xi'an, China

[4] Nebraska Center for Materials and Nanoscience, University of Nebraska, Lincoln, Nebraska 68588, USA

## Abstract

High perpendicular magnetic anisotropy (PMA), a property needed for nanoscale spintronic applications, is rare in oxide conductors. We report the observation of a PMA up to 0.23 $MJ/m^3$ in modestly strained (–0.3%) epitaxial $NiCo_2O_4$ (NCO) films which are room-temperature ferrimagnetic conductors. Spin-lattice coupling manifested as magnetoelastic effect was found as the origin of the PMA. The in-plane $x^2$-$y^2$ states of Co on tetrahedral sites play crucial role in the magnetic anisotropy and spin-lattice coupling with an energy scale of 1 meV/f.u. The elucidation of the microscopic origin paves a way for engineering oxide conductors for PMA using metal/oxygen hybridizations.

Materials and heterostructures of high magnetic anisotropy have been increasingly demanded for energy and information applications. In particular, electrodes with perpendicular magnetic anisotropy (PMA) is needed in nanoscale spintronic devices for their high thermal stability and energy-efficient switching. Most materials or heterostructures of high PMA are based on intermetallic compounds [1–5], multilayers [6,7], or metal/oxide interfaces [8], often with high-cost elements such as Au and Pt. In contrast, transition-metal oxide conductors, despite their advantage of low-cost and structural and chemical stabilities, have rarely been reported to exhibit high PMA.

High magnetic anisotropy originates from structural anisotropy and spin-orbit coupling. In ordered intermetallic compounds containing strongly spin-orbit coupled nonmagnetic (NM) metals (e.g. Pd, Au, and Pt) and ferromagnetic (FM) metals (e.g. Fe and Co), anisotropic crystal structures lead to anisotropic hybridization between the states in NM and FM and consequently high magnetic anisotropy ($\approx$ 5 $MJ/m^3$) [1–5,9]. The structural anisotropy can also be introduced by stacking NM and FM layers for high PMA ($\approx$ 1 $MJ/m^3$) [7]. On the other hand, remarkable PMA has been demonstrated in Co/Ni multilayers ($\approx$ 0.5 $MJ/m^3$) [6] and FM/oxide interfaces ($\approx$ 0.2 $MJ/m^3$) [8,10,11], without the need of the strongly spin-orbit coupled NM. Here, the electronic degeneracy and occupancy are adjusted such that, the 3d states in FM with a large orbital angular momentum (in-plane states) determine the magnetic anisotropy. In particular, at the FM/oxide interface, the 3d electronic states are tuned via the hybridization with oxygen states; this suggests the possibility of having transition-metal oxide with high magnetic anisotropy.

In 3d transition-metal oxides, the hybridization of metal 3d and oxygen 2p states generates a crystal-field splitting $\Delta \sim 1$ eV; the spin-orbit coupling ($\xi \sim 50$ meV) couples these split states and modifies the energy by $\approx <L_z>\xi^2/\Delta$, where $<L_z>$ is the average angular momentum projection along the out-of-plane direction. This energy modification, which gives rise to magnetic anisotropy energy, could reach ~1 meV if $<L_z> \sim 1$; this is why $CoFe_2O_4$, an insulator, indeed exhibits large magnetic anisotropy (~ 1 $MJ/m^3$) and strong spin-lattice coupling, which can be employed to realize PMA in strained films [10,12–22]. For oxide conductors, however, room temperature ferromagnetism is already rare, not to mention high magnetic anisotropy. Widely studied FM oxide conductors, such as $La_{0.7}Sr_{0.3}MnO_3$, unfortunately have low magnetic anisotropy and weak spin lattice coupling due to the dominant $z^2$ state which has a low orbital angular momentum [23].

Inverse spinel $NiCo_2O_4$ (NCO) has recently been demonstrated to be conducting due to the mixed valence [24,25], and ferrimagnetic [24,26–28] from the antialignment of Ni and Co moments above room temperature. In this work, we demonstrate that NCO has a remarkable magnetic anisotropy and spin-lattice coupling which can be employed to generate PMA up to 0.23 $MJ/m^3$ with a –0.3% epitaxial strain [Fig. 1]. Analysis of the microscopic origin of the magnetic anisotropy and the spin-lattice coupling based on single-ion anisotropy reveals the key role of the $x^2$-$y^2$ states in Co atoms on the tetrahedra sites.

Pulsed laser deposition (PLD) was employed to grow epitaxial NCO films between 15 and 20 nm on (001), (110), and (111)-oriented nonmagnetic $MgAl_2O_4$ (MAO) substrates (a = 8.089 Å for MAO, 8.114 Å for NCO, in-plane strain $e_{in} = –0.3\%$). Film growth was conducted with 20 mTorr $O_2$ pressure, 360 ˚C substrate temperature, 5 cm target-to-substrate distance, using a KrF excimer laser (248 nm, 10 Hz, and 2.5 J/cm$^2$). Post-growth annealing was carried out *ex-situ* in 1 atmosphere $O_2$ at 500 ˚C. The crystallinity, thickness, and lattice constants of the films were measured using x-ray diffraction (XRD) with a Rigaku D/Max-B x-ray diffractometer (λ = 1.789 Å) and a Rigaku SmartLab x-ray diffractometer (λ = 1.54 Å). The dependences of magnetization on temperature and magnetic field were measured in a Quantum Design MPMS system. In-plane magnetic anisotropy of NCO (111) films was studied using a home-built Magneto-optical Kerr Effect (MOKE) system at room temperature in a longitudinal configuration with a rotational sample stage [see Supplementary Materials].

X-ray diffraction shows an epitaxial growth of NCO (001) film on MAO (001) substrates with no observable impurity phases [Fig. 2(a)]. The Laue oscillations around the (004) peak suggest a high film quality and a thickness of 17 nm for the displayed film. Reciprocal space mapping [Fig. 2(b)] indicate that the film is fully strained since the in-plane lattice constants of the NCO film coincide with that of the MAO substrate. The NCO peak positions in Fig. 2(a) and that in Fig. 2(b) indicate an out-of-plane strain $e_{\text{out}}$ = 1.3%. For the (110) and (111) oriented films of similar thickness, $e_{\text{out}}$ is 0.8% and 0.6%, respectively [see Supplementary Materials].

Temperature dependence of the magnetization (M-T) measured while cooling in a 100 Oe out-of-plane magnetic field shows an upturn [Fig. 2(c)], indicating a transition to magnetic ordering at $T_C$ = 323 K. Figure 2(d) shows a clear square magnetization-field (M-H) hysteresis loops at 150 K, with an out-of-plane magnetic field (along the [001] axis); the coercivity is 450 Oe and the remnant magnetization is 94% of the saturation value. In contrast, the M-H relation has no significant remanence and no measurable coercivity with an in-plane field (along the [100] or [010] axes); the saturation field is 20 kOe. The distinct in-plane and out-of-plane M-H relations reveal PMA in the NCO (001) films, where [001] is the easy axis. By comparing the M-H relations with in-plane and out-of-plane magnetic fields, one can extract the anisotropy energy $K_u$ for the PMA of NCO (001) [Fig. 1].

The PMA in the NCO (001) films, can be understood as a result of spin-lattice coupling and the broken cubic symmetry due to the biaxial epitaxial strain. On a phenomenological level, spin-lattice coupling can be described as magnetoelastic effect with the Landau theory using free energy

$$F = K_1(\alpha_1^2\alpha_2^2 + \alpha_2^2\alpha_3^2 + \alpha_3^2\alpha_1^2) + B_1\left(\alpha_1^2 e_{xx} + \alpha_2^2 e_{yy} + \alpha_3^2 e_{zz}\right) + B_2(\alpha_1\alpha_2 e_{xy} + \alpha_2\alpha_3 e_{yz} + \alpha_3\alpha_1 e_{zx}) \quad (1),$$

where $K_1$ is the magnetic anisotropy constant, $\alpha_1$, $\alpha_2$, and $\alpha_3$ are the directional cosines of the magnetization with respect to *x*, *y*, and *z* axes respectively ($\Sigma\alpha_i^2 = 1$), $e_{ij}$ are components of the strain tensor, $B_1$ and $B_2$ are the longitudinal and shear magnetoelastic coupling constants respectively. The first term in Eq. (1) corresponds to the magnetic anisotropy of cubic symmetry, while the second and third terms described the magnetoelastic coupling. Without strain ($e_{ij} = 0$), if $K_1 > 0$, by minimizing $F$, one finds that the global easy axes are [100], [010], and [001], which are equivalent under the cubic symmetry [see Supplementary]; the global hard axes are along [111] axis or its equivalent. These results can be visualized in Fig. 2(e), where the easy (hard) axes correspond to the energy minima (maxima).

For the NCO/MAO (001) films (nonzero strain: $e_{xx} = e_{yy} = e_{in} < 0$, $e_{zz} = e_{out} > 0$), the free energy along the [100] and [001] axis are $F_{[100]} = B_1 e_{xx}$ and $F_{[001]} = B_1 e_{zz}$, respectively; the observed easy axis along the [001] direction [Fig. 2(d)] requires $F_{[001]} < F_{[100]}$ or $B_1 < 0$, as illustrated in Fig. 2(f).

To fully characterize the magnetic anisotropy and the magnetoelastic effect, we also studied the *M-H* relations for the (111) and (110)-oriented NCO films.

The *M-H* relation of the NCO (111) films with in-plane and out-of-plane field directions, all show an "S" shaped loop with a small remanence (12%-15%) and coercivity (700 Oe along [111]; 500 Oe along other directions) [Fig. 3(a)], indicating that they are not easy axes. The easy axis is most likely tilted with non-zero projections in both in-plane and out-of-plane directions. To investigate the tilted easy axes, we carried out MOKE measurement using the longitudinal mode, which measures the projection of magnetization in the direction of the reflected light. Using this method, one may observe a normal or an inverse *M-H* hysteresis loop, when the in-plane azimuthal angle between the easy axis and the reflected light are less than or greater than $\pi/2$, respectively (see Supplementary Materials). As shown in Fig. 3(b), both normal and inverse *M-H* loops were observed when the film was rotated about the [111] axis. Using a negative coercivity to distinguish the inverse *M-H* loops from the normal ones, the in-plane anisotropy can be visualized using the polar plot of the coercivity [Fig. 3(b)], where the larger coercivity means closer to the easy axis. A triangular symmetry is revealed, and the in-plane projection of the easy axis appears to be along the [11-2] (or equivalent) directions. The [100], [010], and [001] directions satisfy the geometric symmetry for the easy axes observed in Fig. 3(b).

For the NCO/MAO (111) films, the nonzero strain is $e_{xy} = e_{yz} = e_{zx} = -e_{in} > 0$. According to Eq. (1), a positive (negative) $B_2$ suggests that the compressive strain increases (decreases) the energy of the [111] axis. Experimentally, the measured hysteresis along the out-of-plane direction [111] and the in-plane directions $[11\bar{2}]$ are similar [Fig. 3(a)], suggesting that the energy of the [111] direction is reduced from the global maximum, indicating that $B_2 < 0$. The free energies of the [100], [010], and [001] axes are not affected since all the longitudinal strains ($e_{xx}$, $e_{yy}$, and $e_{zz}$) are zero. For small strain, the global easy axes remain close to these directions, consistent with the MOKE observation [see Supplementary Materials].

The *M-H* relation of the NCO (110) films exhibits a slightly canted shape with a coercivity of 550 Oe and remnant magnetization 91% of saturation magnetization when the magnetic field is along the [100] in-plane direction. In contrast, when the magnetic field is along the in-plane [1-10] and out-of-plane [110] directions, the *M-H* relation has a minimal hysteresis with a saturation field 10 kOe. For the NCO/MAO (110) films, the nonzero strain is $e_{xx} = e_{yy} = (e_{in} + e_{out})/2 > 0$, $e_{zz} = e_{in} < 0$, and $e_{xy} = (e_{out} - e_{in})/2 > 0$. The free energies of the [100] and [010] axes reduce, while that of [001] axis increases, because $B_1 < 0$. Therefore, [001] becomes a local easy axis, consistent with slightly canted *M-H* loop measured when the field is along the in-plane [001] axis, while the global easy axes remain close to the [100] and [010] directions [see Supplementary Materials].

One may determine $K_1$, $B_1$, and $B_2$ from the magnetic anisotropy energy extracted from the *M-H* relations for the NCO films of different orientations. The results are listed in Table I, as well as in

Fig. 1. The tunability of the magnetic anisotropy is highlighted by the large magnetoelastic coupling coefficients $B_1$ and $B_2$.

Next, we analyze the microscopic origin of the spin-lattice coupling in terms of the effect of strain on the single-ion magnetic anisotropy energy via the spin-orbit coupling. We employ a model Hamiltonian using a one-electron picture

$$H = \sum_i \left[\frac{p_i^2}{2m} - \frac{Ze}{4\pi\varepsilon_0 r_i} + V_{CF} + \xi \vec{S}_i \cdot \vec{l}_i + E_{ex}\vec{S}_i \cdot \hat{B}_{ex}\right],$$

where $p_i$, $l_i$, $S_i$, $\vec{r}_i$ are momentum, orbital angular momentum, spin, and position vector of the $i$th electron, $-\frac{Ze}{4\pi\epsilon_0 r_i}$ and $V_{CF}(\vec{r}_i)$ are the potential energy due to the ion core and the crystal field respectively, $\hat{B}_{ex}$ is an exchange field that generates the energy gap $E_{ex}$ between spins of opposite directions, $e$, $m$, $\varepsilon_0$, g, $\hbar$ are the electronic charge, electronic mass, vacuum permittivity, Landé g-factor, and reduced Planck constant. The spin-lattice coupling can be understood as that the strain modifies the electronic orbital states by changing the local environment of the magnetic ions ($V_{CF}$), followed by the change of their preferred spin orientations due to the spin-orbit coupling.

In the unit cell of NCO, eight low-spin $Ni^{3-\delta}$ ions and eight high-spin $Co^{2+\delta}$ ions are in $NiO_6$ octahedra [Figs. 4(a), $O_h$ symmetry] and $CoO_4$ tetrahedra [Fig. 4(b), $T_d$ symmetry] respectively, where $\delta < 1$ which indicates the mixed valences [24,25]. The other eight $CoO_6$ octahedra are contain low-spin $Co^{3+}$, which do not contribute to magnetism. The Co and Ni 3d states are split into doubly degenerate $e_g$ states and triply degenerate $t_{2g}$ states due to the corresponding $V_{CF}$. Under the biaxial compressive strain which reduces the cubic symmetry to tetragonal, these states further split [Fig. 4(a) and 4(b)].

We simulate the crystal field by replacing the oxygen atoms with point charges in $NiO_6$ and $CoO_4$. The total energy on a magnetic ion $E_t$ is calculated by summing the energy of the individual electrons [6] according to the population in Figs. 4(a) and (b), where $\delta$=0.5 is assumed. The single-ion magnetic anisotropy manifests in the dependence of $E_t$ on the direction of $\hat{B}_{ex}$. As an example, for in the (001) NCO films, the single-ion magnetic anisotropy is defined as $E_{SIMA} = E_{t,x} - E_{t,z}$, where $E_{t,x}$ and $E_{t,z}$ are $E_t$ when $\hat{B}_{ex}$ is along the $x$ (in-plane) and $z$ (out-of-plane) axes respectively. The epitaxial strain $\Delta a/a$, where $a$ is the bulk lattice constant, is introduced by distorting the $NiO_6$ and $CoO_4$ local environment according to the lattice constant change, which are $\Delta a$ and $-2\Delta a$ for in-plane and out-of-plane axes, respectively. The simulated $E_{SIMA}$ as a function of strain is shown in Fig. 4(c). For both $Ni^{3-\delta}$ and $Co^{2+\delta}$, under the compressive strain ($\Delta a < 0$) which generates a tetragonal distortion, $E_{SIMA}$ is positive, suggesting that the $c$ axis (out-of-plane direction) is the easy axis, which is consistent with the experimental observation.

To reveal more microscopic detail of the effect of strain on magnetic anisotropy, here we analyze $Co^{2+\delta}O_4$ tetrahedra as an example since it shows a larger effect in Fig. 4(c). In this case, the 3d electronic configuration can be viewed as a half-filled shell plus an electron in the $|x^2\text{-}y^2\rangle$ state and a fractional occupation in the $z^2$ state, due to the tetragonal distortion that generates an $S_4$ symmetry, as shown in Fig. 4(b) and 4(d). Since the half-filled shell is not expected to contribute to the magnetic anisotropy, the electron in the $|x^2\text{-}y^2\rangle$ state dominates the anisotropy. As illustrated in Fig. 4(d), if the spin is along the *z* axis, the $|x^2-y^2, S_z=1/2>$ state couples to the $|xy, S_z=1/2>$ state to lower its energy with a coupling strength $<x^2-y^2, S_z=1/2|\, \xi \vec{S}_i \cdot \vec{l}_i |\, xy, S_z=1/2> = \xi$. On the other hand, when the spin is along the *x* axis, the $|x^2-y^2, S_x=1/2>$ state couples to the $|xz, S_x=1/2>$ state to lower its energy with a coupling strength $<x^2-y^2, S_x=1/2|\, \xi \vec{S}_i \cdot \vec{l}_i |\, xz, S_x=1/2> = \xi/2$, which is smaller than that when the spin is along the *c* axis. Therefore, the compressive strain results in an out-of-plane magnetic anisotropy. Hence, the 3d $|\, x^2-y^2>$ state of Co in the $Co^{2+\delta}O_4$, plays a key role in the spin-lattice coupling of NCO due to its potentially large orbital angular momentum along the *z* axis. Assuming the magnitude of $\xi$, $V_{CF}$, and $E_{ex}$ as 0.05, 1 and 5 eV respectively, the single-ion magnetic anisotropy is found to be ~1 meV per formula unit, as show in Fig. 4(c); this translates to ~1 $MJ/m^3$ in magnetic anisotropy energy, in fair agreement with the observed values in Table I.

In conclusion, we have demonstrated a remarkable PMA in the (001)-oriented NCO/MAO epitaxial films above room temperature which can be understood as a result of the spin-lattice coupling manifested as magnetoelastic effect. The microscopic origin of spin-lattice coupling has been explained using the effect of strain on the single-ion magnetic anisotropy energy due to spin-orbit coupling. The demonstration and elucidation of the strong tunability of magnetic anisotropy in NCO, indicate the possibility of high PMA in oxide conductors. This adds material structures, such as NCO/MAO/NCO tunnel junction of enhanced magnetoresistance [29] into nanoscale spintronic devices. In addition, it opens up another route toward electrical and mechanical control of magnetism above room temperature.

**Acknowledgments**

This project was primarily supported by the National Science Foundation (NSF), DMR under award DMR-1454618. C.S. acknowledges the support by the Nebraska Center for Energy Sciences Research on magnetic imaging. J.W. acknowledges the support of REU program of Nebraska Center for Materials and Nanoscience on magneto optical studies. The research was performed in part in the Nebraska Nanoscale Facility: National Nanotechnology Coordinated Infrastructure, and the Nebraska Center for Materials and Nanoscience, which are supported by the National Science Foundation under award ECCS: 1542182, and the Nebraska Research Initiative.

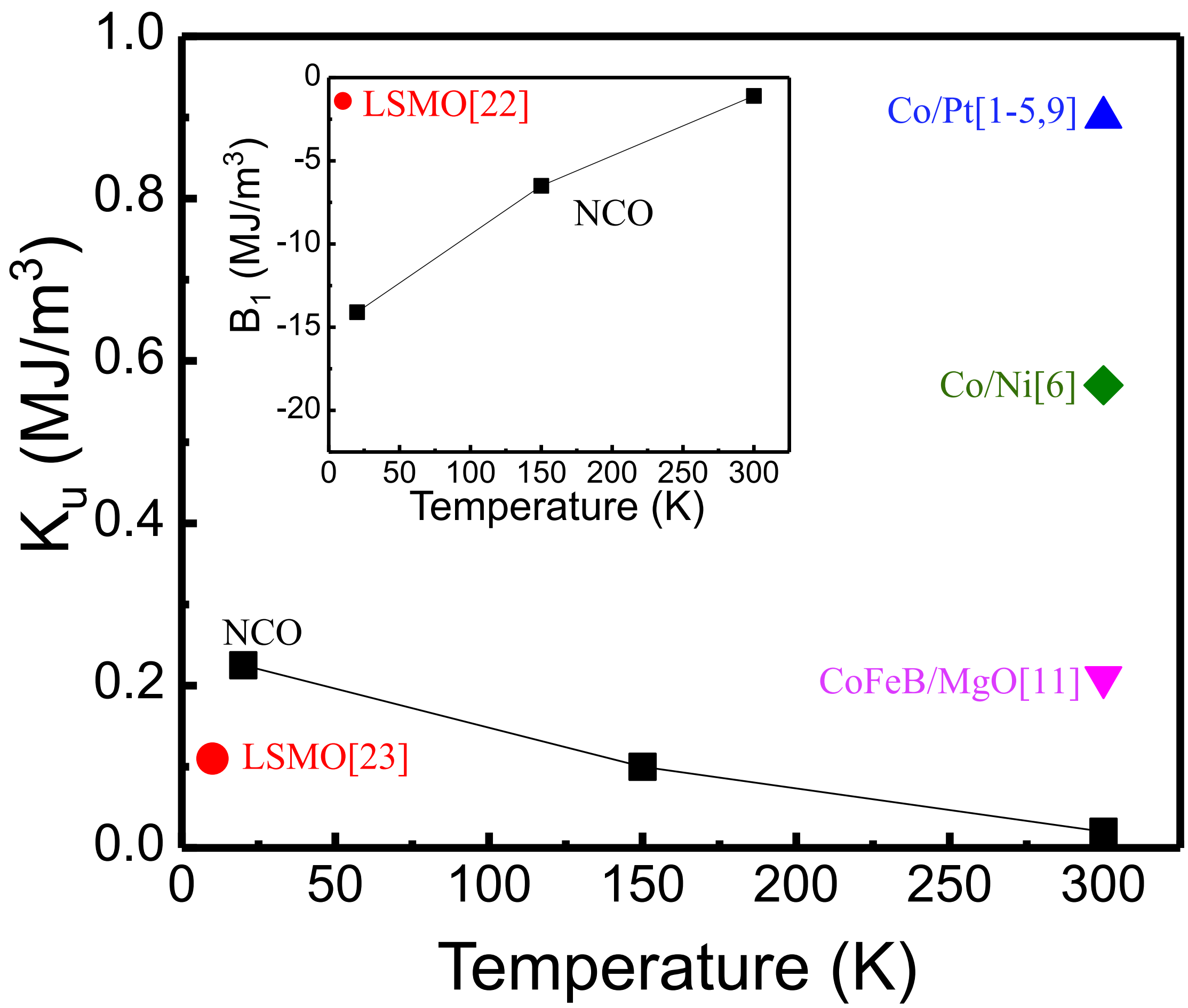

**Figure 1**. Uniaxial magnetic anisotropy energy $K_u$ of strained $NiCo_2O_4$ (–0.3%) and $La_{0.7}Sr_{0.3}MnO_3$ (–2.1%) films, Co/Pt and Co/Ni multilayers, and CoFeB/MgO heterostructures. Inset: magnetoelastic coupling coefficient $B_1$ of $NiCo_2O_4$ and $La_{0.7}Sr_{0.3}MnO_3$.

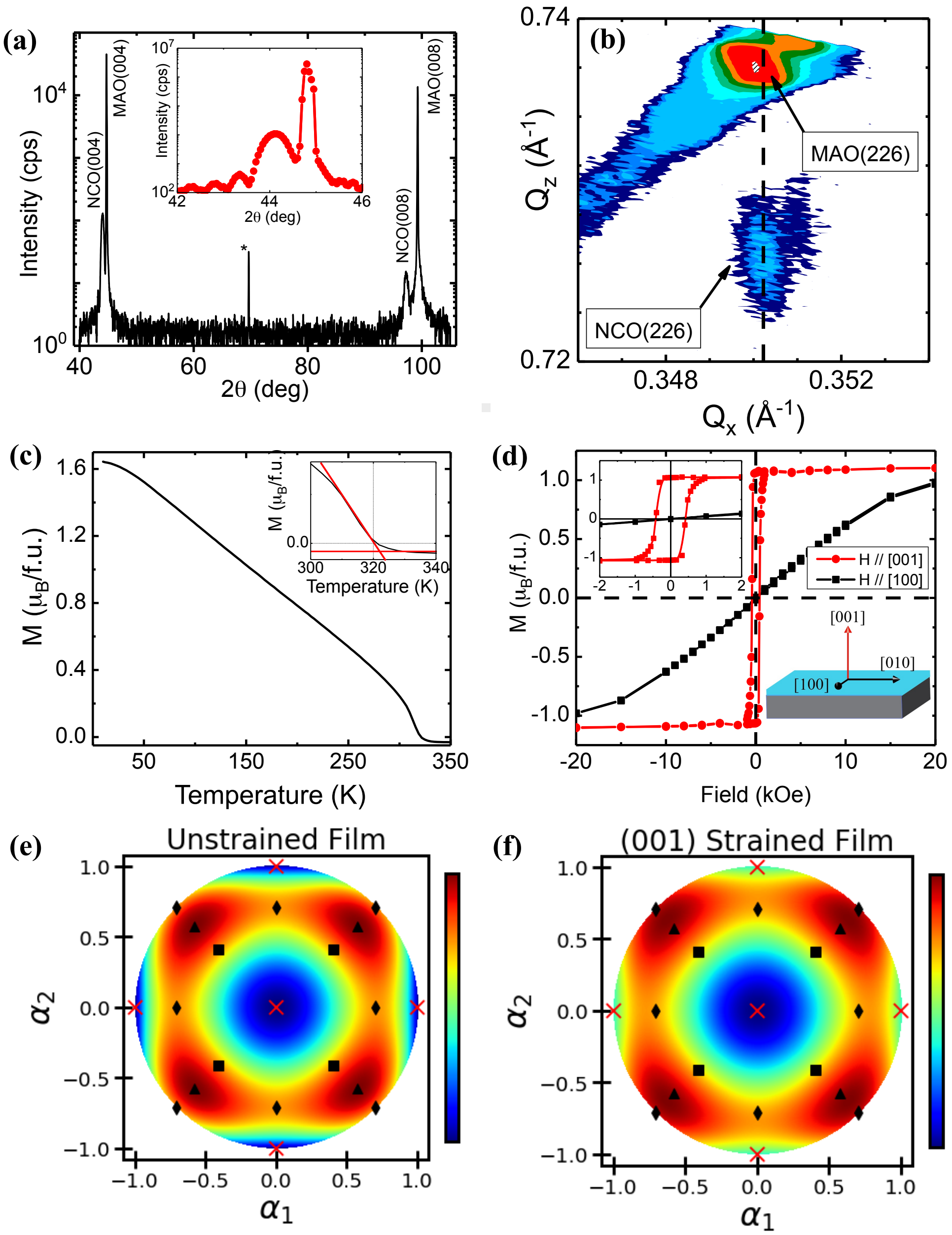

**Figure 2. Structural and magnetic behavior of (001)-oriented films. (a)** θ-2θ scan of NCO/MAO(001). Inset is a scan of the NCO (004) peak. **(b)** Reciprocal space mapping of (226) peaks of NCO and MAO. Alignment of peaks along the $Q_x$ axis indicates in-plane lattice matching of the film to the MAO substrate. **(c)** *M-T* curve of NCO/MAO(001) film field-cooled in a 100 Oe out-of-plane magnetic field. The inset shows the upturn of magnetization occurring at $T_C = 323$ K. **(d)** *M-H* relations at 150 K. The upper-left inset shows magnetization behavior closer to the origin. Lower-right inset shows a sketch of the sample. The magnetic anisotropy energy $F(\alpha_1, \alpha_2)$ of a cubic material without strain **(e)**, and with a compressive biaxial strain in the (001) plane **(f)**, is calculated from the Landau theory. The cross, diamond, square, and triangle symbols indicate [100], [110], [11-2], [111] directions or their cubic equivalent, respectively. The calculation uses the experimentally-determined values of constants $B_1/K_1 = -11.96$ and $B_2/K_1 = -18.45$.

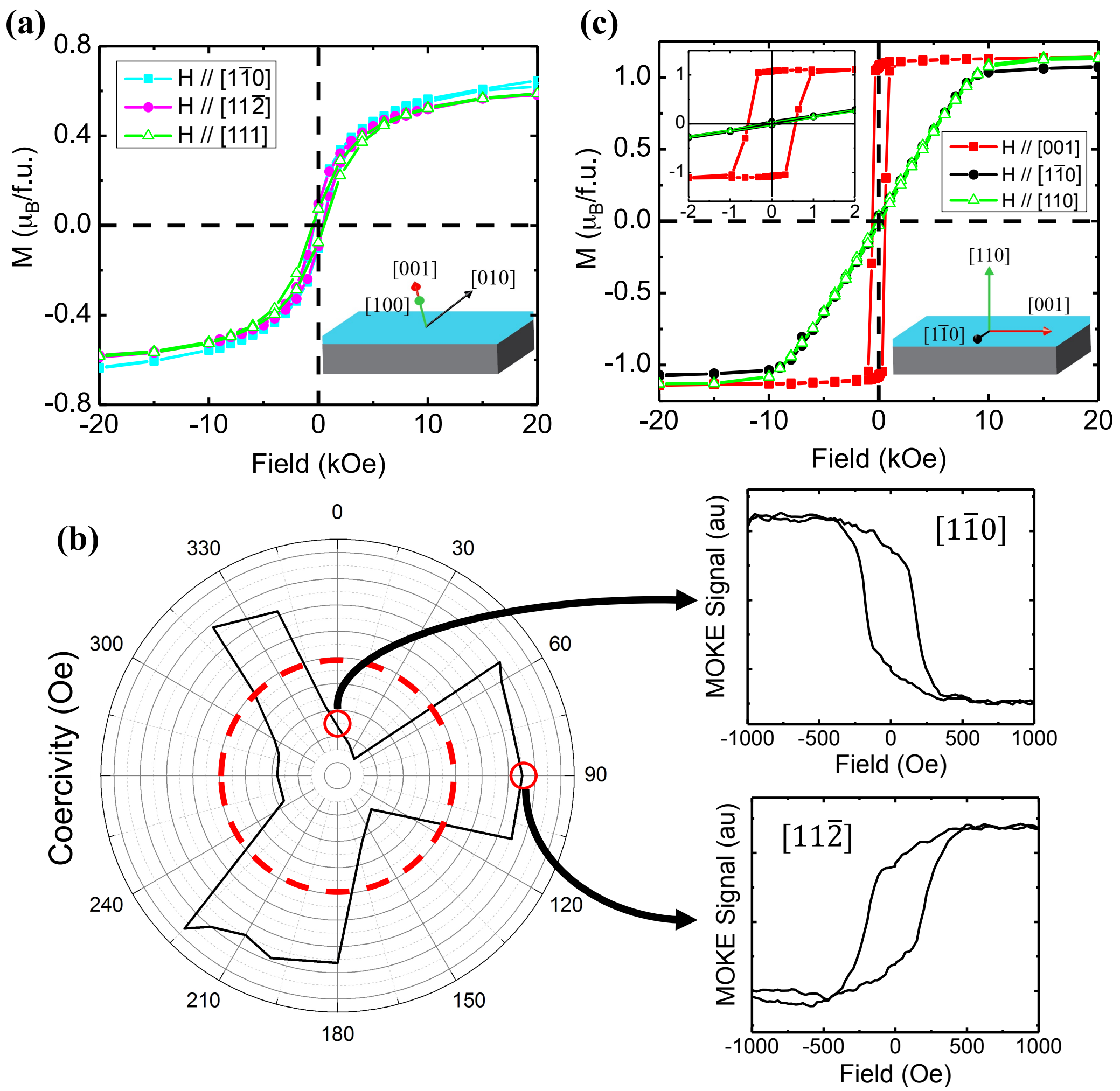

**Figure 3. Magnetic behavior of (111)- and (110)-oriented NCO films. (a)** *M-H* relation along the out-of-plane and two in-plane directions of the (111)-oriented films measure at 150 K. Inset is a schematic of the (111)-oriented film. **(b)** Polar plot of the coercivities of the hysteresis loops measured using MOKE at room temperature (see text). The red dashed circle indicates zero coercivity; i.e. values outside the circle have a hysteresis loop with positive saturation at high fields ("normal"), while values inside the circle have a hysteresis loop with positive saturation at negative fields ("inverted"). Examples of each type of loop are shown at the angles. The corresponding crystallographic direction along which the magnetic field is applied is shown on the individual plots. **(c)** *M-H* relation at 150 K for the (110)-oriented films. Upper-left inset further shows magnetization behavior closer to the origin. Lower-right inset shows a sketch of the sample with axes labeled.

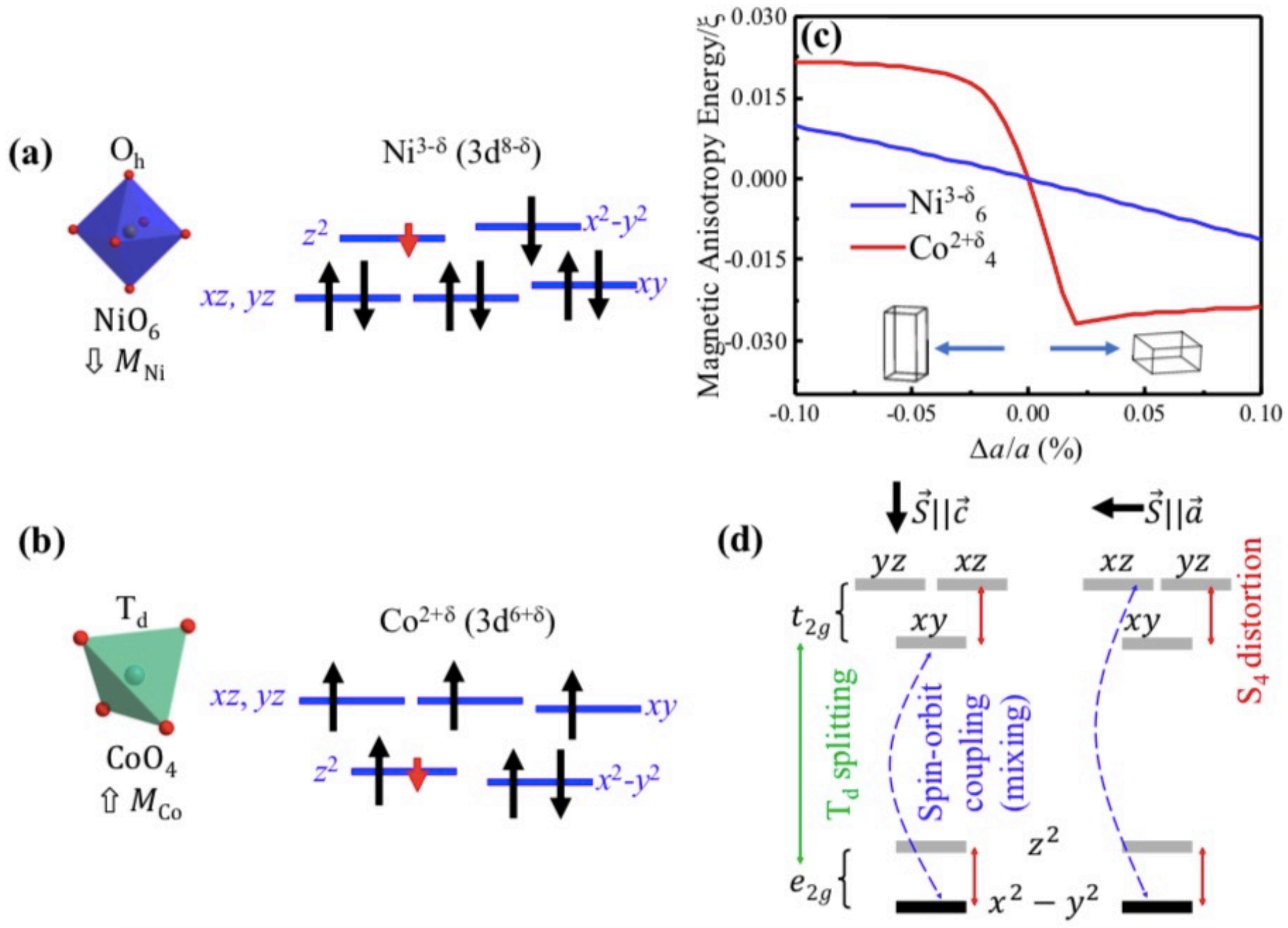

**Figure 4. Microscopic model of the effect of the biaxial strain in the (001) plane. (a)-(b)** Octahedral and tetrahedral environments of magnetic nickel and cobalt sites, respectively. The energy diagrams show the splitting between $e_g$ and $t_{2g}$ levels, as well as the smaller splitting within the symmetry groups due to the tetragonal strain distortions. The short arrows represent partial (δ) occupation of the orbital. **(c)** The magnetic anisotropy energy calculated from the single-ion magnetic anisotropy as a function of in-plane biaxial strain, where $\xi = 50$ meV is the spin-orbit coupling constant; the magnitude of $V_{CF}$ and $E_x$ are set as 1 eV and 5 eV respectively. **(d)** Relative energies of mixing states due to the spin-orbit coupling. The energy gain is larger when the spin is along the *c* axis than that when the spin is perpendicular to the *c* axis, leading to magnetocrystalline anisotropy.

**Table I.** Magnetic anisotropy and magnetoelastic coupling coefficient of $NiCo_2O_4$ measured in this work.

| $K_u$ (MJ/m$^3$) | $K_1$ (MJ/m$^3$) | $B_1$ (MJ/m$^3$) | $B_2$ (MJ/m$^3$) |
|---|---|---|---|
| 0.23 (20 K)<br>0.1 (150 K)<br>0.02 (300 K) | 0.54 (150 K)<br>0.08 (300 K) | –14.1 (20 K)<br>–6.5 (150 K)<br>–1.1 (300 K) | –10 (150 K)<br>–1.5 (300 K) |

**Supplementary Information**

# Perpendicular Magnetic Anisotropy in Conducting $NiCo_2O_4$ Films from Spin-Lattice Coupling

Corbyn Mellinger[1], Jace Waybright1[1,2], Xiaozhe Zhang[1,3], Caleb Schmidt[1], and Xiaoshan Xu[1,4]

[1] Department of Physics and Astronomy, University of Nebraska, Lincoln, Nebraska 68588, USA
[2] Department of Physics, South Dakota State University, Brookings, SD 57007 USA
[3] School of Materials Science and Engineering, Xi'an Polytechnic University, Xi'an, 710048, China
[4] Nebraska Center for Materials and Nanoscience, University of Nebraska, Lincoln, Nebraska 68588, USA

## S1. SQUID & XRD Supplementary

An expanded view of the M-H curves from the main text (Figs. 2(b), 3(a), 3(c)) is given in Figure S1(a)-(c). Applied field of up to 50 kOe confirm the films have reached their full saturation magnetizations. Interestingly, the (001) oriented film in Fig. S1(a) has differing saturation magnetizations for the out-of-plane [001] and in-plane [100] directions. This phenomenon has been seen in cobalt ferrite thin films, and can be attributed to the coupling of the orbital angular momentum with the spin-orbit interaction, which arise due to the strain-induced recombination of d-orbitals [1].

## S2. MOKE Supplementary

Magneto-optical Kerr Effect (MOKE) measurements on the NCO/MAO(111) structures are made at room temperature using an experimental configuration as outlined in Sato [2]. An intensity-stabilized HeNe laser ($\lambda$ = 632 nm) is used as the probing laser. The first polarizer generates the s-polarization of light for interaction with the sample. Magnetic fields of up to 5 kOe are generated using an electromagnet. Additionally, the samples are mounted on an Aluminum stand, which is attached to a goniometer with markings every 1˚. The surface of the Al stand was machined to be flat, to avoid tilting of the sample during rotation. In the longitudinal MOKE geometry, the magnetic field is applied in the plane of the sample, and the polarized laser incident at ~30˚ from the sample normal. A diagram of the sample holder is shown in Figure S2(a). A field sweep at a fixed angle is done several times and averaged to reduce noise. The goniometer is then rotated, and the field sweeps are repeating, giving hysteresis loops at several angles. Importantly, since the magnetic field is not rotated along with the sample, the crystallographic direction along which the field is applied changes, and the measurement looks at the magnetization along that direction, allowing for in-plane magnetic anisotropy of the sample to be studied. Through the rotation, hysteresis loops would change from so-called "regular" loops (in which application of a positive saturation field brings the Kerr rotation to its maximum) to so-called "inverted" loops (in which application of a positive saturation field brings Kerr rotation to its minimum). We analyze these loops using the coercivity, denoting the "regular" loops with coercivity positive and equal to its magnitude, and "inverted" loops with coercivity negative and equal to its magnitude. Raw data from the angular-dependent hysteresis loops are show in Figure S2(b), with the lines drawn to highlight the three-fold symmetry seen. These graphs correspond to the coercivities used in Figure 3(b).

Our observation of angle-dependent hysteresis leads directly from the strong anisotropy of the NCO along the ⟨001⟩ axes. Typical explanations of Kerr effect measurements assume sample magnetization aligning with the external field, but this is not necessarily the case. This can be seen by considering a 3-dimensional Stoner-Wohlfarth model of magnetic domains, in which the magnetic energy has terms from anisotropy and Zeeman contributions: $E = K\sin^2(\alpha) - HM\sin(\theta)\cos(\phi) = K\left[\sin^2(\alpha) - \frac{HM}{K}\sin(\theta)\cos(\phi)\right]$, with $\theta, \phi$ the spherical angles of magnetization, and $\alpha$ the angle between magnetization and anisotropy axis. As the anisotropy scale $K$ increases, the magnetization direction of minimum energy tilts further from the field direction and more towards the anisotropy axis.

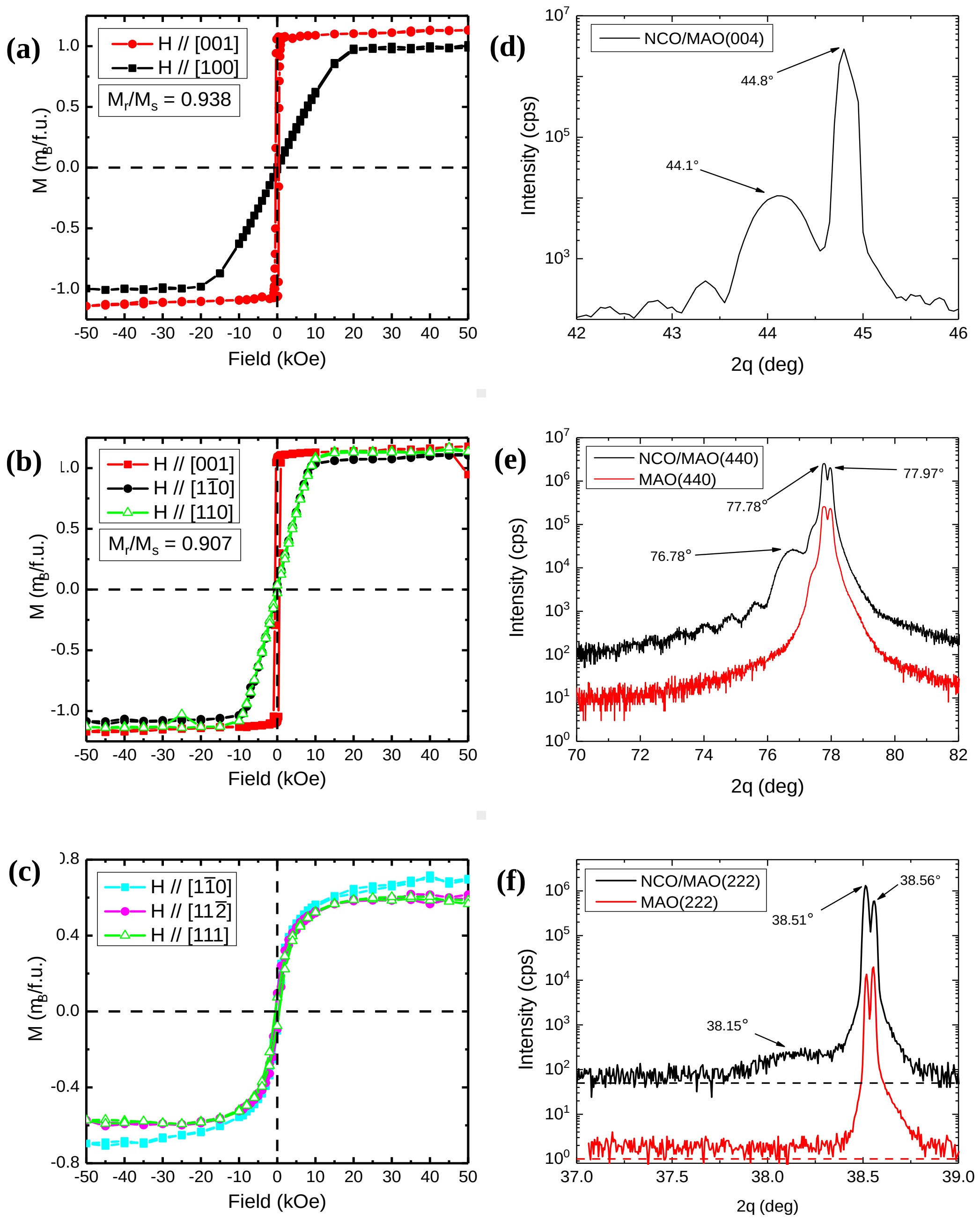

**Figure S1 | (a)-(c):** Expansion of SQUID figures from Figs. 2(a)-(b), 3(a) to 50 kOe. **(d)-(f):** Out-of-plane diffraction data for the three orientations of film studied. 2θ values of the peaks are labeled for clarity. Double-peaks appear for the substrate due to the presence of significant Cu-K$\alpha_2$ in addition to Cu-K$\alpha_1$. The determination of out-of-plane strain for the three sample orientations were determined using θ-2θ XRD on the Rigaku SmartLab, and the plots are given in **(d)-(f),** with **(e)** and **(f)** also showing the diffraction of patterns of the substrates alone. All x-ray data uses Cu-Kα wavelengths from a Rigaku SmartLab system.

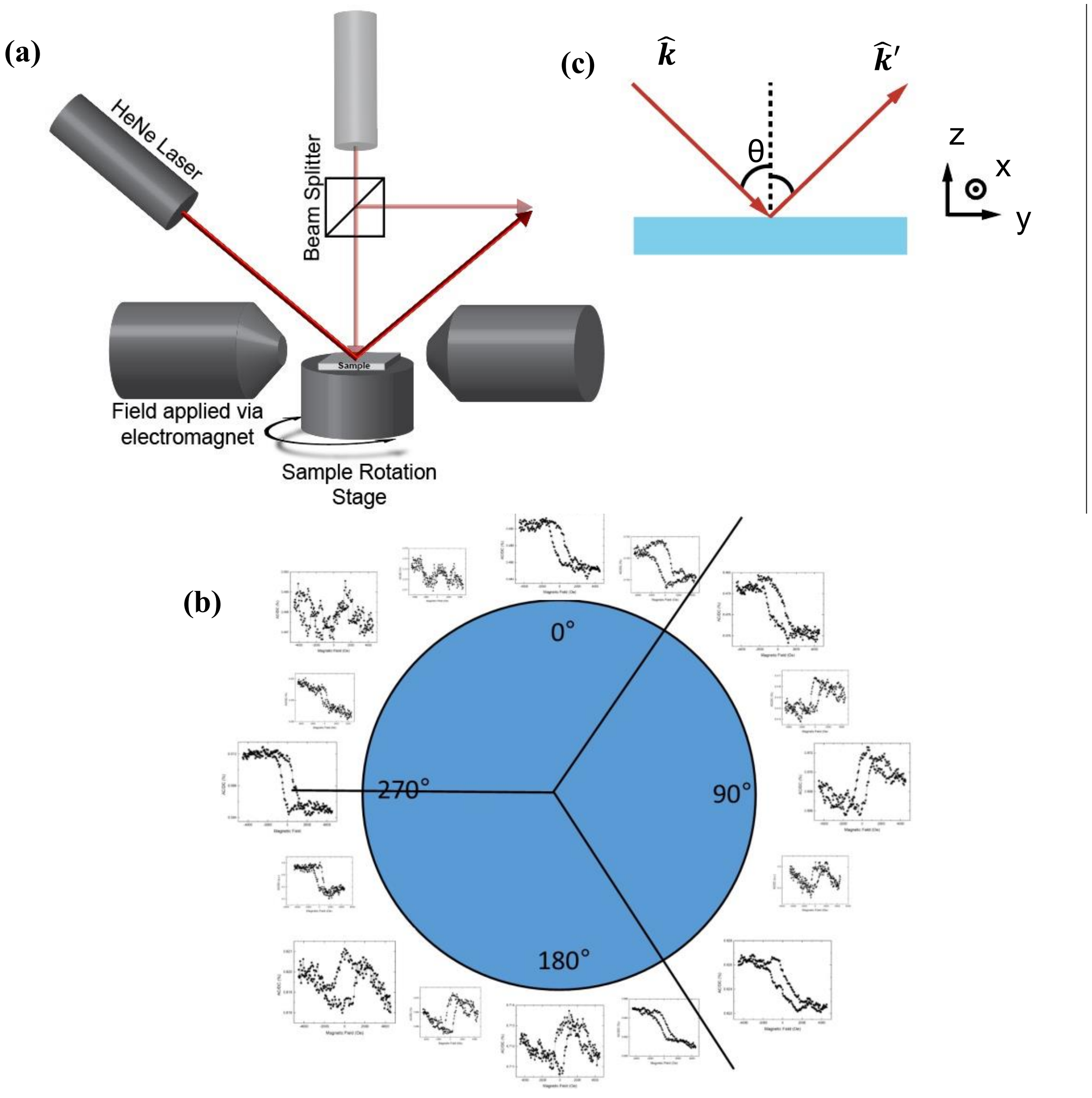

**Figure S2 | (a)** Diagram of the relative positions of sample, rotation stage, electromagnet, and laser used in longitudinal MOKE measurements. Transparent portion of the figure is the configuration used shows the configuration used when measuring polar MOKE using in-plane magnetic field, a so-called "hybrid MOKE" configuration. Since the magnetization lies largely out-of-plane, this measurement provides non-zero Kerr rotation. **(b)** Example of raw data from a longitudinal MOKE measurement as a function of rotation angle. Data is more noisy at points where the magnetization is much lower along the plane of reflection. **(c)** Diagram of the geometry used in the Lorentz force derivation just above.

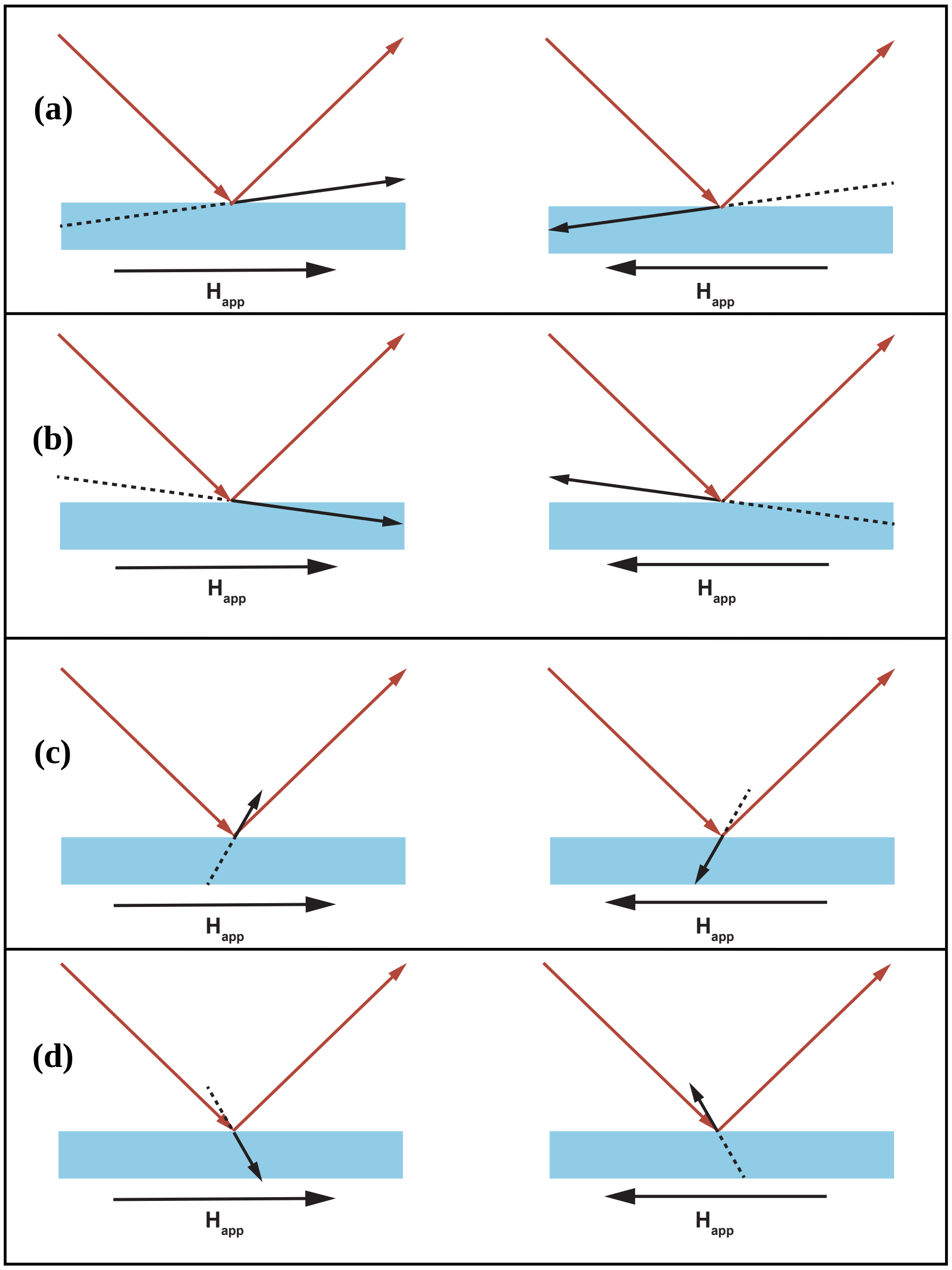

**Figure S3 | Visualization of the MOKE measurement under several relative orientations of $\hat{\boldsymbol{k}}'$, $\hat{\boldsymbol{m}}$.** In both **(a)** and **(b)**, the magnetization lies nearly in-plane. In **(a)**, $\hat{k}' \cdot \hat{m} > 0$ for $\vec{H}$ pointing right, and $< 0$ when pointing left. **(b)** shows the same sample and magnetization orientation, but rotated by 180°. In this case, again, $\hat{k}' \cdot \hat{m} > 0$ for field pointing right, and $< 0$ for field pointing left. By contrast, **(c)** and **(d)** have a magnetization lying much closer to perpendicular to the sample surface. In **(c)**, $\hat{k}' \cdot \hat{m} > 0$ for applied field pointing right, and $< 0$ for field pointing left. However, once the sample is rotated by 180° as shown in **(d)**, $\hat{k}' \cdot \hat{m} < 0$ for field pointing right, and $> 0$ for field pointing left. The signs of the Kerr rotation have not changed between **(a)** and **(b)**, while the signs of the Kerr rotation have changed between **(c)** and **(d)**. These latter two figures apply to the highly out-of-plane easy axes for the NCO/MAO (111) structure, leading to the flipping of hysteresis loops as the sample is rotated.

A consequence of the relative orientations of the laser incidence and magnetization is that a large enough misalignment of the directions leads to a flipping of the sign of the magneto-optic effect. Explicitly, $\hat{k}' \cdot \hat{m} = 0$ when the angle of incidence satisfies $\tan(\theta) = \frac{m_z}{m_y}$, and on either side of this value, the effect of the magnetization switches signs. This is displayed visually in the supplementary Fig S3(a)-(d). The severe misalignment of the laser k'-vector and magnetization direction seen in the NCO/MAO(111) measurements leads directly to the emergence of inverted hysteresis loops, and the three-fold rotational symmetry of the angle-dependent MOKE measurement.

We can further analyze the effect of magnetization direction relative to incidence direction modeling the Magneto-optic Kerr Effect in terms of the Lorentz force experienced by charges in the magnetized material. In our case of s-polarized light (i.e. polarization perpendicular to the plane of reflection), the electric field forces an oscillation of the charges, which then experience a Lorentz force. With light polarization along the x-axis (i.e. s-polarized light), charges will be forced to oscillate along this same axis. Given the most general magnetization direction $\hat{m} = (m_x, m_y, m_z)$, the force experienced will be $\hat{F} \sim \hat{x} \times (m_x\hat{x} + m_y\hat{y} + m_z\hat{z}) = m_y\hat{z} - m_z\hat{y}$. The induced p-polarization is what will be measured, and the axis of the p-polarization lies within the plane of reflection, and perpendicular to the direction of reflected light. For a general direction of incident light $\hat{k} = \sin(\theta)\,\hat{y} - \cos(\theta)\,\hat{z}$, as illustrated in Figure S3(c), the reflected light will follow $\hat{k}' = \sin(\theta)\,\hat{y} + \cos(\theta)\,\hat{z}$. The rotated polarization will have component along the p-polarization direction, which can be represented as $\hat{p} = -\cos(\theta)\,\hat{y} + \sin(\theta)\,\hat{z}$. The measured Kerr rotation will be proportional to the amount of new polarization lying in this plane, or $\hat{F} \cdot \hat{p} = m_z\cos(\theta) + m_y\sin(\theta) = \hat{k}' \cdot \hat{m}$. Repeating calculations for the case of p-polarized light yields the similar result that the Lorentz force induces rotation proportional to $\hat{k} \cdot \hat{m}$.

## S3. Magnetic Force Microscopy

Direct imaging of the domain structure of NCO on MAO (001), (110), and (111) substrates is performed using magnetic force microscopy (MFM) in a Bruker ScanAsyst system, with a commercially available magnetic tip. A bar magnet with a strength of ~400 Oe was used to pole the sample before scanning. With the magnetic tip normal to the sample, magnetization pointing normal to the surface is sensitive to scanning, while in-plane magnetization is undetectable. Thus, the MFM provides a confirmation of easy of out-of-plane versus in-plane magnetization. To differentiate the magnetic effects of tip deflection from topographical features, an additional scan using AFM is done in the area of interest, and the AFM component is removed from the total scan signal. Typical MFM and AFM scans of (001), (110), and (111) oriented samples are shown in Figs. S4 (a)-(f). No domains are seen in the (110) and (111) samples, while domains of ~500 nm are seen in (001) samples, demonstrating magnetization out-of-plane for this orientation.

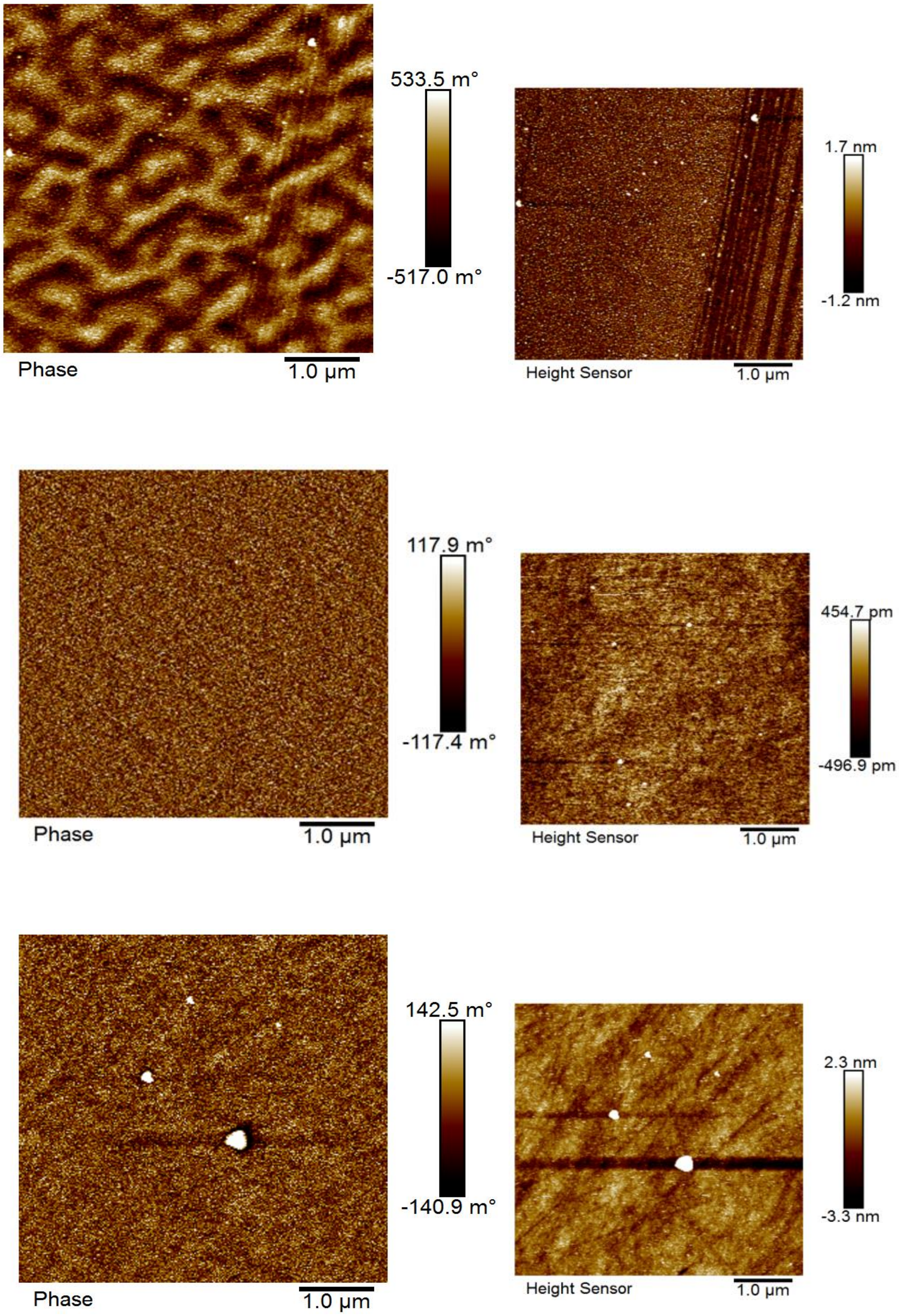

**Figure S4 | Magnetic force microscopy and atomic force microscopy data for the three NCO orientations studied. (a)**: NCO/MAO (001) MFM; **(b)**: NCO/MAO (110) MFM; **(c)**: NCO/MAO (111) MFM. **(d)-(f)** are the AFM images taken over the regions shown in **(a)-(c)**, respectively. Only **(a)** shows considerable magnetic behavior in the out-of-plane direction measured, which is consistent with the measurements by SQUID showing only the (001)-oriented NCO having easy magnetization out-of-plane.

## S4. Landau Theory of Spin-Lattice Coupling

The full magnetic cubic anisotropy energy and magnetoelastic energy is given as: $F = K_1(\alpha_1^2\alpha_2^2 + \alpha_2^2\alpha_3^2 + \alpha_3^2\alpha_1^2) + B_1(\alpha_1^2 e_{xx} + \alpha_2^2 e_{yy} + \alpha_3^2 e_{zz}) + B_2(\alpha_1\alpha_2 e_{xy} + \alpha_2\alpha_3 e_{yz} + \alpha_3\alpha_1 e_{zx})$, with terms as described in the main text. Explicitly in the case of strained (001) films, the compressive normal stress on the cubic faces of the unit cell makes off-diagonal shear strain terms zero, with $e_{xx}, e_{yy} < 0$ and $e_{zz} > 0$. Compressive strain in the (001) plane means $e_{xx} = e_{yy} \equiv e_{in}$; $e_{zz} = e_{out}$, where $e_{in}$ and $e_{out}$ are the in-plane and out-of-plane strains. In all cases, $e_{in} = -0.003$ from the lattice mismatch and the observation that films are epitaxially strained to the substrate (Figure 1(b) in the main text). The value of $e_{out}$ is determined experimentally from θ-2θ measurements shown in Figure S1(d)-(f). Additionally, the squares of directional cosines sum to unity so that $\alpha_3$ can be expressed in terms of $\alpha_1, \alpha_2$. Applying these simplifications:

$F = K_1(\alpha_1^2 + \alpha_2^2 - \alpha_1^4 - \alpha_2^4 - \alpha_1^2\alpha_2^2) + B_1(e_{in} - e_{out})(\alpha_1^2 + \alpha_2^2) + B_1 e_{out}$. First derivatives with respect to $\alpha_1, \alpha_2$ are taken and set to zero to find potential critical values of energy, and the sign of second derivatives with respect to $\alpha_1, \alpha_2$ show the possible magnetization directions to be maxima, minima, or saddle points in energy. For the (001) samples, the procedure gives solutions summarized below in Table S1.

**Table S1: Solutions to free-energy minimization of strained (001)-oriented NCO. $\langle 110\rangle$ and $\langle 111\rangle$ directions remain difficult to magnetize regardless of relation between $K_1$, $B_1$. $\Delta e \equiv e_{in} - e_{out}$.**

| # | $\alpha_1$ | $\alpha_2$ | $\frac{\partial^2 F}{\partial\alpha_1^2}$ | $\frac{\partial^2 F}{\partial\alpha_2^2}$ | Axis | $K_1 + \Delta e B_1 > 0$ | $K_1 + \Delta e B_1 < 0$ |
|---|---|---|---|---|---|---|---|
| **1** | 0 | 0 | $2K_1 + 2\Delta e B_1$ | $2K_1 + 2\Delta e B_1$ | $[0,0,\pm 1]$ | Easy | Hard |
| **2** | 0 | $\pm\sqrt{\frac{K_1 + \Delta e B_1}{2K_1}}$ | $K_1 + \Delta e B_1$ | $-4(K_1 + \Delta e B_1)$ | $[0, \pm 1, \pm 1 + \delta]$ | Saddle | Saddle |
| **3** | $\pm\sqrt{\frac{K_1 + \Delta e B_1}{2K_1}}$ | 0 | $-4(K_1 + \Delta e B_1)$ | $K_1 + \Delta e B_1$ | $[\pm 1, 0, \pm 1 + \delta]$ | Saddle | Saddle |
| **4** | $\pm\sqrt{\frac{K_1 + \Delta e B_1}{3K_1}}$ | $\pm\sqrt{\frac{K_1 + \Delta e B_1}{3K_1}}$ | $-\frac{8(K_1 + \Delta e B_1)}{3}$ | $-\frac{8(K_1 + \Delta e B_1)}{3}$ | $[\pm 1, \pm 1, \pm 1 + \delta]$ | Hard | Easy |

For example, Solution #4 has $\alpha_1^2 = \alpha_2^2$ so the axis will lie in $(\pm 1, \pm 1, 0)$ plane. Since $\Delta e B_1$ is small, $\alpha_3^2$ will be very close to $\frac{1}{3}$ (and thus the to the $\langle 111\rangle$ axes), but will in general differ, giving the small rotation, denoted generally by δ in Table S1.

Anisotropy energy can be directly measured from hysteresis curves, which then provides values of anisotropy and magnetoelastic constants. The (001) samples were measured with magnetization along [100] and [001], as shown in Fig. 2(a), which have magnetic energies $e_{in}B_1$ and $e_{out}B_1$, respectively. In other words, the cubic magnetic anisotropy is broken, with [100] and [010] axes raising in energy and [001] axis lowering.

The (111) NCO experiences only shear strains, with stresses along the body diagonals of the unit cell, in other words, the strain tensor satisfies $e_{ii} = 0; e_{ij} = e > 0$. The magnetic energy involves only $K_1$ and $B_2$ constants: $F_{(111)} = K_1(\alpha_1^2\alpha_2^2 + \alpha_2^2\alpha_3^2 + \alpha_3^2\alpha_1^2) - e_{in}B_2(\alpha_1\alpha_2 + \alpha_2\alpha_3 + \alpha_3\alpha_1)$. Magnetic energies are tabulated in Table S3.

(110) oriented NCO has a combination of both shear and normal strains, which means $K_1, B_1$, and $B_2$ terms are all included in the magnetic energy expression. Explicitly written in terms of $e_{out}$ and $e_{in}$: $F_{(110)} = K_1(\alpha_1^2\alpha_2^2 + \alpha_2^2\alpha_3^2 + \alpha_3^2\alpha_1^2) + B_1\left(\frac{e_{out}+e_{in}}{2}\right)(\alpha_1^2 + \alpha_2^2) + e_{in}B_1\alpha_3^2 + B_2\left(\frac{e_{out}-e_{in}}{2}\right)\alpha_1\alpha_2$. Energies along measured directions are listed in Table S4.

**Table S2: Summary of magnetization energetics along measured directions for the (001)-oriented NCO in strained and unstrained cases.**

| Mag. Direction | F(unstrained) | F(strained) | $F_{(001)}^{[100]} - F_{(001)}^{[001]}$ |
|---|---|---|---|
| $[001]$ | 0 | $e_{out}B_1$ | $(e_{in} - e_{out})B_1$ |
| $[100]$ | 0 | $e_{in}B_1$ | |

**Table S3: Summary of magnetization energetics along measured directions for the (111)-oriented NCO in strained and unstrained cases.**

| Mag. Direction | F(unstrained) | F(strained) | $F_{(111)}^{[111]} - F_{(111)}^{[11\bar{2}]}$ |
|---|---|---|---|
| $[111]$ | $\frac{K_1}{3}$ | $\frac{K_1}{3} - e_{in}B_2$ | $\frac{K_1}{12} - \frac{3e_{in}B_2}{2}$ |
| $[11\bar{2}]$ | $\frac{K_1}{4}$ | $\frac{K_1}{4} + \frac{e_{in}B_2}{2}$ | |
| $[1\bar{1}0]$ | $\frac{K_1}{4}$ | $\frac{K_1}{4} + \frac{e_{in}B_2}{2}$ | —— |

**Table S4: Summary of magnetization energetics along measured directions for the (110)-oriented NCO in strained and unstrained cases.**

| Mag. Direction | F(unstrained) | F(strained) | $F_{(110)}^{[1\bar{1}0]} - F_{(110)}^{[110]}$ |
|---|---|---|---|
| $[001]$ | 0 | $e_{in}B_1$ | —— |
| $[110]$ | $\frac{K_1}{4}$ | $\frac{K_1}{4} + B_1\left(\frac{e_{out}+e_{in}}{2}\right) - \frac{B_2}{2}\left(\frac{e_{out}-e_{in}}{2}\right)$ | $B_2\left(\frac{e_{out}-e_{in}}{2}\right)$ |
| $[1\bar{1}0]$ | $\frac{K_1}{4}$ | $\frac{K_1}{4} + B_1\left(\frac{e_{out}+e_{in}}{2}\right) + \frac{B_2}{2}\left(\frac{e_{out}-e_{in}}{2}\right)$ | |

With all 3 orientations of NCO, the constants $K_1, B_1, B_2$ can be determined experimentally. From Tables S2-S4, we see that (001) *M-H* curves can uniquely determine the value of $B_1$, that $B_2$ can be determined based on (110) *M-H* curves, and that the (111) *M-H* curves can determine the relation between $B_2$ and $K_1$, and thus give $K_1$. In the special observed case that $F_{(111)}^{[111]} - F_{(111)}^{[11\bar{2}]} = 0$, the relation between $K_1$ and $B_2$ is $K_1 = 18e_{in}B_2$. The anisotropy energy is directly measurable from the hysteresis curves, using the saturation magnetization and field required to reach saturation: $U = \frac{1}{2}\mu_0 M_s H_A$. Given these relations, we can find the magnetization energy constants which are listed in Table I of the main text: $K_1 = 5.4 \times 10^5 \frac{J}{m^3}, B_1 = -6.5 \times 10^6 \frac{J}{m^3}, B_2 = -1.0 \times 10^7 \frac{J}{m^3}$.

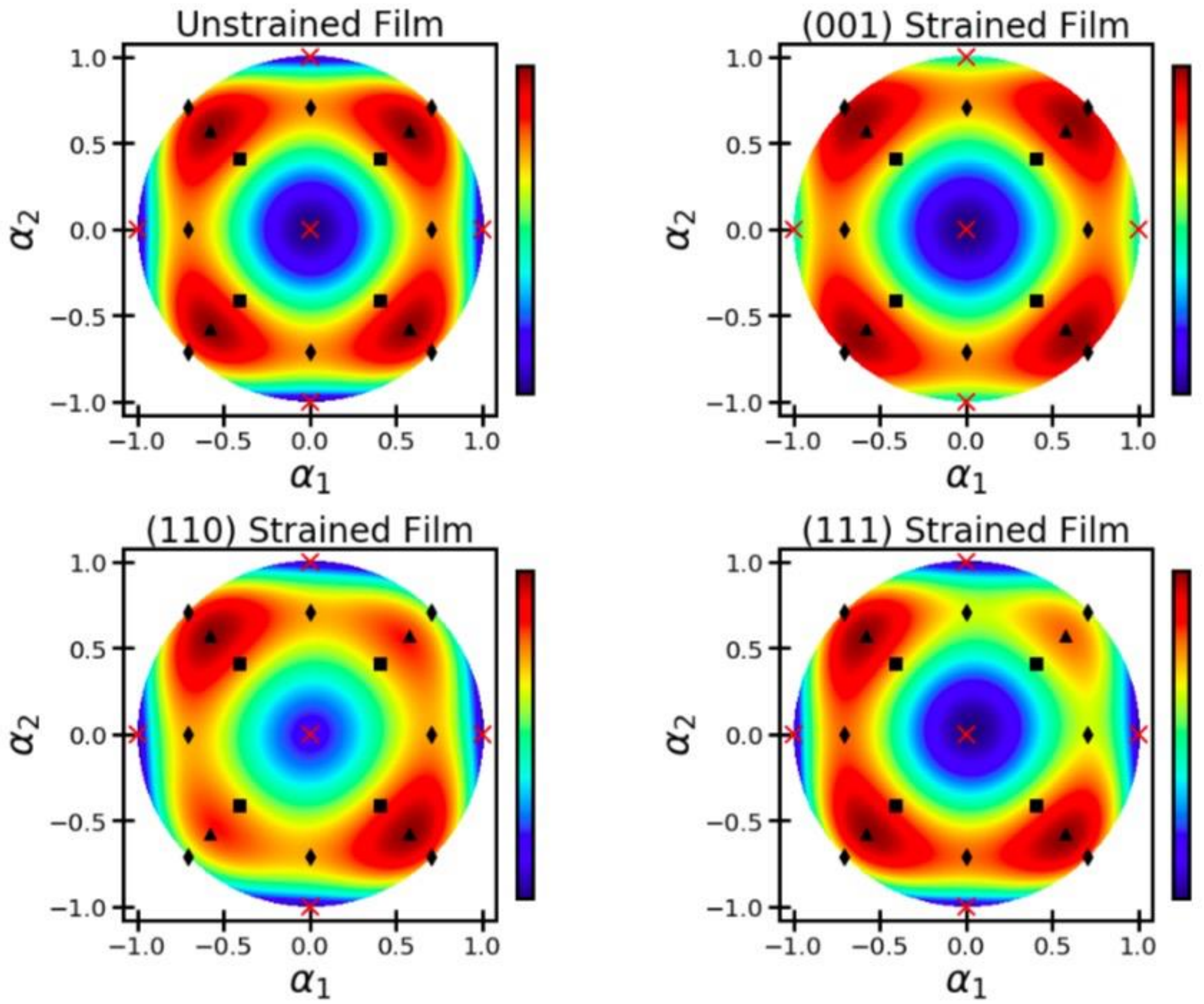

**Figure S5 | Summary of magnetization energies from Landau theory considerations. (a):** Unstrained cubic magnetic anisotropy. The [001], [100], and [010] magnetization directions are degenerate global minima of magnetization energy. **(b):** Cubic material under biaxial strain in (001) plane. The [001] axis has lowered relative to the other edge directions, giving the out-of-plane anisotropy. Panes **(a), (b)** are exactly the same as those in Fig. 2(c) and (d), respectively. **(c):** Magnetic energy of cubic material under biaxial strain in (110) plane. The [001] axis remains as a global minimum, but the true magnetization energy minima lie along [100], [010] axes. **(d):** Magnetic energy of cubic material under biaxial strain in (111) plane. The magnetization energy degeneracy along the $\langle 111 \rangle$ axes is broken, with the true out-of-plane [111] direction easier to magnetize than others in that class. As in Figure 2, the cross, diamond, square, and triangle markers indicate the $\langle 001 \rangle, \langle 110 \rangle, \langle 11\bar{2} \rangle$, and $\langle 111 \rangle$ crystal directions, respectively.

Relative energies can be determined for general magnetization direction by plotting the magnetic free energy as a function of magnetization directions $\alpha_1, \alpha_2$ with experimentally-determined strain and anisotropy energy values used as given in Table I. The unstrained plot in Fig. S5(a) shows the cubic anisotropy, i.e. lower energies along [001] ($\alpha_1 = \alpha_2 = 0$), [100] ($\alpha_1 = 1, \alpha_2 = 0$), and [010] ($\alpha_1 = 0, \alpha_2 = 1$) axes. Energy landscape for the three orientations of strained films are shown, using strain values as determined by θ-2θ x-ray diffraction [Figs. S1 (d)-(f)]. RSM of the (001)-oriented film at the (226) film plane [Fig. 1(b)] confirms in-plane lattice matching, so pinned strains of -0.3% are assumed for all in-plane directions, with out-of-plane strains determined by x-ray diffraction. For ease of determination of common axis directions, the diamonds are $\langle 110 \rangle_c$ axes, triangles are $\langle 111 \rangle_c$ axes, squares are $\langle 11\bar{2} \rangle_c$ axes, and red crosses are $\langle 001 \rangle_c$ axes.

The true easy magnetization direction for the strained (110) film was not measured in our study; the true easy axes [100] and [010] lie neither exactly in-plane or out-of-plane of the sample. The strain is not sufficient to bring the [001] axis out of a local minimum, so it appears relatively easy. Considering that the other two axes measured are [110] and [1-10], the in-plane [001] axis is the easiest measured.

In the case of the (111) strained film, the $\langle 001 \rangle$ axes remain the global minima, but some deviation from this direction is noticeable; the dark blue minimum along [001] shifts slightly towards the [111] direction. Additionally, the [111] axis is lower in energy than the [-111], [1-11], and [11-1] axes. It is clear that the effect of the strain is to make the out-of-plane direction easier or harder to magnetize.

## S5. Microscopic Mechanism of Spin-lattice Coupling

Modeling the octahedral and tetrahedral environments of Ni and Co atoms is done by considering the Hamiltonian involving the free energy of the electron, electric field generated by the negatively-charged oxygen sites, spin-orbit coupling, and exchange interaction, explicitly written as $H = \sum_i \left[\frac{p_i^2}{2m} - \frac{Ze}{4\pi\varepsilon_0 r_i} + V_{CF} + \xi \vec{S}_i \cdot \vec{l}_i + E_x \vec{S}_i \cdot \hat{B}_{ex}\right]$. The Hamiltonian is then diagonalized, and eigenstates corresponding to the $d_{x^2-y^2}, d_{z^2}, d_{xy}, d_{yz}, d_{xz}$ orbitals are obtained, and their eigen-energies determined. These yield splitting of $e_g$ and $t_{2g}$ symmetry's energies as determined from classical crystal field theory. Then, perturbations of the octahedra and tetrahedra are analyzed using perturbation theory, and their energy levels determined. As an illustration of the physics involved, one such example is done analytically in the following sections, but full analysis is performed using numerical methods and plotting using open-access pysci and matplotlib packages.

### I: Crystal-field, Structural distortion, Exchange field components

Real wavefunctions $d_{x^2-y^2}, d_{z^2},$ etc. are linear combinations of the spin-orbit basis, and the separation between higher-energy $e_g$ states (in the octahedral environment) represented by an additional energy term D, with D ~ exchange splitting energy.

The crystal field energy consists of Coulomb interaction between electrons and the charged oxygen ligands. Since the energy scale is relatively small in comparison with the ionic interactions, the energy can be expand in terms of Legendre polynomials, using the notation of Arfken (7$^{th}$ ed.,pg. 737):

$$V_{CF}(\vec{r}) = \sum_{i}^{N} \frac{Ze^2}{4\pi\varepsilon|\vec{R}_i - \vec{r}|} = \frac{Ze^2}{4\pi\varepsilon}\sum_{i}\sum_{k=0}^{\infty}\frac{1}{r_>}\left(\frac{r_<}{r_>}\right)^k P_k(\cos(\omega_i))$$

with $r_<, r_>$ being the smaller and larger of $\vec{R}_i$ and $\vec{r}$, Z being the charge of the $N$ ligands, and $i$ running over the positions of the crystal field ligands. The Legendre polynomials may be expanded in terms of the spherical harmonics:

$$P_k(\cos(\omega_i)) = \frac{4\pi}{2k+1}\sum_{m=-k}^{k} Y_{km}(\theta,\phi)\, Y_{km}^{*}(\theta_i,\phi_i)$$

,where $(R_i, \theta_i, \phi_i)$ and $(r, \theta, \phi)$ are the spherical coordinates of the ligands and of the electron, respectively. Within the localized single-electron treatment, we can safely replace $r_<$ with $r$ and $r_>$ with $R_i$, giving an energy term with the form

$$V_{CF} = \frac{Ze^2}{4\pi\varepsilon}\sum_{i}^{N}\sum_{k=0}^{\infty}\sum_{m=-k}^{k}\frac{1}{R_i}\left(\frac{r}{R_i}\right)^k \frac{4\pi}{2k+1} Y_{km}(\theta,\phi) Y_{km}^{*}(\theta_i,\phi_i)$$

For a general configuration of crystal field ligands. In the case of the $O_h$ symmetry, all $|R_i| = R$, we recast the equation in a simpler form:

$$V_{CF}(\vec{r}) = \frac{Ze^2}{4\pi\varepsilon R}\sum_{k=0}^{\infty}\sum_{m=-k}^{k}\left(\frac{r}{R}\right)^k \sqrt{\frac{4\pi}{2k+1}}\gamma_{km}\ Y_{km}(\theta,\phi)$$

,where $\gamma_{km} \equiv \sqrt{\frac{4\pi}{2k+1}}\sum_{i=1}^{N}\left(\frac{R}{R_i}\right)^{k+1} Y_{km}^{*}(\theta_i,\phi_i)$, with R being the median distance from metal ion to oxygen ligand. The purpose of recasting the equation in this form will become apparent later in the derivation.

Matrix elements are populated using the inner product of $V_{CF}$ with the spatial wavefunctions, which in terms of the separable hydrogen electron wavefunctions:

$$V_{CF}^{m_1,m_2} = \iiint R_{n,l}^{*}\, Y_{l,m_1}^{*} V_{CF}(\vec{r}) R_{n,l} Y_{l,m_2} r^2 \sin(\theta)\, dr\, d\theta\, d\phi \qquad (\boldsymbol{S1})$$

$$= \sum_{k=0}^{\infty}\sqrt{\frac{4\pi}{2k+1}}\gamma_{k,m_1-m_2}\, U_{n,l,k} C^k(l,m_1,l,m_2)$$

Where $U_{n,l,k} \equiv \frac{Ze^2}{4\pi\varepsilon R^{k+1}}\int_0^{\infty} R_{nl}^2\, r^{k+2} dr$, $C^k(l,m_1,l,m_2) \equiv \sqrt{\frac{4\pi}{2k+1}}\iint Y_{l,m_1}^{*} Y_{k,m_1-m_2} Y_{l,m_2} \sin(\theta)\, d\theta\, d\phi$, the latter of which are known as the Gaunt coefficients and are related to the Clebsch-Gordan coefficients. Due to the rules of the CG coefficients, only k from 0 to $l_1 + l_2$ are nonzero, restricting the summation to a finite number of terms. Further, due to the symmetry of the $O_h$ system, only even values of k will possibly contribute to the matrix element. In the case of the $l_1 = l_2 = 2$ for d-electrons this leaves only $k = \{0,2,4\}$. Additionally, by inspection one can verify that $\gamma_{2,0} = 0$ for the octahedral crystal field configuration.

Finally, based on the octahedral symmetry in which the 6 oxygens have distance R and the angles of the 6 ligands are $(\theta,\phi) = (0,0), (\pi,0), \left(\frac{\pi}{2},\frac{\pi}{2}\right), \left(\frac{\pi}{2},\frac{3\pi}{2}\right), \left(\frac{\pi}{2},\pi\right), \left(\frac{\pi}{2},0\right)$, the sum over $N = 6$ reduces to:

$$H_{CF} = U_{3,2,4}\begin{bmatrix} \gamma_{4,0}C^4_{-2,-2} & 0 & 0 & 0 & \gamma_{4,4}C^4_{-2,2} \\ 0 & \gamma_{4,0}C^4_{-1,-1} & 0 & 0 & 0 \\ 0 & 0 & \gamma_{4,0}C^4_{0,0} & 0 & 0 \\ 0 & 0 & 0 & \gamma_{4,0}C^4_{1,1} & 0 \\ \gamma_{4,4}C^4_{2,-2} & 0 & 0 & 0 & \gamma_{4,0}C^4_{2,2} \end{bmatrix}$$

where $l_1 = l_2 = 2$ is suppressed from the Gaunt coefficients, which are now written $C^k_{m_1,m_2}$. Also note that $U_{n,l,k}$ provides an energy scale which may be factored out for the purpose of determining relative energy splittings, though some approximation may be made through the evaluation of the appropriate integral in its definition. Through calculation or a table of each coefficient, we get $C^4_{-2,-2} = C^4_{2,2} = \frac{1}{21}, C^4_{-1,-1} = C^4_{1,1} = -\frac{4}{21}, C^4_{0,0} = \frac{6}{21}, C^4_{-2,2} = C^4_{2,-2} = \frac{\sqrt{70}}{21}, \gamma_{4,4} = \gamma_{4,-4} = \sqrt{\frac{35}{8}}, \gamma_{4,0} = \frac{7}{2}$, and thus

$$H^{(1)}_{CF} = \frac{U_{3,2,4}}{6}\begin{bmatrix} 1 & 0 & 0 & 0 & 5 \\ 0 & -4 & 0 & 0 & 0 \\ 0 & 0 & 6 & 0 & 0 \\ 0 & 0 & 0 & -4 & 0 \\ 5 & 0 & 0 & 0 & 1 \end{bmatrix}$$

Diagonalization of this matrix to determine energy eigenvalues gives the splitting of $e_g$, $t_{2g}$ symmetric levels.

The effect of compressive strain on the octahedron is to move oxygens in the xy-plane to distance $R - a$ from the metal ion, and move the polar oxygens to distance $R + c$ from the center, where a, c are small strains which are unrelated to the lattice constants of the crystal. In this case, $\gamma_{2,0} \neq 0$. Calculation gives $\gamma_{2,0} = 2\left\{\left(1 - \frac{3c}{R}\right)^{-3} - \left(1 - \frac{3a}{R}\right)^{-3}\right\} \approx \frac{6(a-c)}{R}$. Changes to the $\gamma_{4,0}, \gamma_{4,4}$ terms are proportional to $\left(\frac{R}{R+a}\right)^5$ and $\left(\frac{R}{R+c}\right)^5$, and neglecting these higher order terms is warranted. Thus, the distortion introduces new terms involving only $\gamma_{4,0}$ along the diagonal of the matrix. Explicitly:

$$H^{(2)}_{CF} = \frac{U_{3,2,2}}{7}\frac{6(a-c)}{R}\begin{bmatrix} -2 & 0 & 0 & 0 & 0 \\ 0 & 1 & 0 & 0 & 0 \\ 0 & 0 & 2 & 0 & 0 \\ 0 & 0 & 0 & 1 & 0 \\ 0 & 0 & 0 & 0 & -2 \end{bmatrix}$$

This interaction is spin-independent, so the full 10x10 matrices for the $|l_z, s_z\rangle$ basis are:

$$H_{CF} = D\begin{bmatrix} 1 & 0 & 0 & 0 & 5 & & & & & \\ 0 & -4 & 0 & 0 & 0 & & & & & \\ 0 & 0 & 6 & 0 & 0 & & & 0 & & \\ 0 & 0 & 0 & -4 & 0 & & & & & \\ 5 & 0 & 0 & 0 & 1 & & & & & \\ & & & & & 1 & 0 & 0 & 0 & 5 \\ & & & & & 0 & -4 & 0 & 0 & 0 \\ & & 0 & & & 0 & 0 & 6 & 0 & 0 \\ & & & & & 0 & 0 & 0 & -4 & 0 \\ & & & & & 5 & 0 & 0 & 0 & 1 \end{bmatrix} + d\begin{bmatrix} -2 & 0 & 0 & 0 & 0 & & & & & \\ 0 & 1 & 0 & 0 & 0 & & & & & \\ 0 & 0 & 2 & 0 & 0 & & & 0 & & \\ 0 & 0 & 0 & 1 & 0 & & & & & \\ 0 & 0 & 0 & 0 & -2 & & & & & \\ & & & & & -2 & 0 & 0 & 0 & 0 \\ & & & & & 0 & 1 & 0 & 0 & 0 \\ & & 0 & & & 0 & 0 & 2 & 0 & 0 \\ & & & & & 0 & 0 & 0 & 1 & 0 \\ & & & & & 0 & 0 & 0 & 0 & -2 \end{bmatrix}$$

where $D = \frac{U_{3,2,4}}{6}$, $d = \frac{U_{3,2,2}}{7}\frac{6(a-c)}{R}$ are the corresponding energy scales of the base and perturbed crystal fields.

Diagonalizing the Hamiltonian, without the spin-orbit contribution, and solving for eigenvalues, eigenvectors yields to first order the $e_g$, $t_{2g}$ orbitals again, with distortion terms $d$ splitting the degeneracy. The wavefunctions are given below in Table 5. Importantly, these wavefunctions are all subject to orbital quenching, in which the total angular momentum of each real wavefunction is zero. Consideration of spin-orbit terms below partially undo this quenching, which will give rise to the magnetocrystalline anisotropy observed in the NCO.

Finally, the effect of exchange interaction is modeled using the tensor product of 5x5 identity matrix with the corresponding x, y, z Pauli spin matrices: $S_{x,y,z} = \sigma_{x,y,z} \otimes I_5$.

**Table S5: Solutions to Hamiltonian, neglecting spin-orbit coupling terms.**

| State | Real Wavefunction | Wavefunction in $\vert L_z, S_z\rangle$ basis | Energy |
|---|---|---|---|
| $\varphi_1$ | $\vert z^2,\uparrow\rangle$ | $\vert 0,\uparrow\rangle$ | $A + D + 2d$ |
| $\varphi_2$ | $\vert xy,\uparrow\rangle$ | $\frac{1}{\sqrt{2}}(\vert 2,\uparrow\rangle - \vert -2,\uparrow\rangle)$ | $A - 2d$ |
| $\varphi_3$ | $\vert x^2-y^2,\uparrow\rangle$ | $\frac{1}{\sqrt{2}}(\vert 2,\uparrow\rangle + \vert -2,\uparrow\rangle)$ | $A + D - 2d$ |
| $\varphi_4$ | $\vert xy,\uparrow\rangle$ | $\frac{1}{\sqrt{2}}(\vert 1,\uparrow\rangle + \vert -1,\uparrow\rangle)$ | $A + d$ |
| $\varphi_5$ | $\vert yz,\uparrow\rangle$ | $\frac{1}{\sqrt{2}}(\vert 1,\uparrow\rangle - \vert -1,\uparrow\rangle)$ | $A + d$ |
| $\varphi_6$ | $\vert z^2,\downarrow\rangle$ | $\vert 0,\downarrow\rangle$ | $D + 2d$ |
| $\varphi_7$ | $\vert xy,\downarrow\rangle$ | $\frac{1}{\sqrt{2}}(\vert 2,\downarrow\rangle - \vert -2,\downarrow\rangle)$ | $-2d$ |
| $\varphi_8$ | $\vert x^2-y^2,\downarrow\rangle$ | $\frac{1}{\sqrt{2}}(\vert 2,\downarrow\rangle + \vert -2,\downarrow\rangle)$ | $D - 2d$ |
| $\varphi_9$ | $\vert xy,\downarrow\rangle$ | $\frac{1}{\sqrt{2}}(\vert 1,\downarrow\rangle + \vert -1,\downarrow\rangle)$ | $d$ |
| $\varphi_{10}$ | $\vert yz,\downarrow\rangle$ | $\frac{1}{\sqrt{2}}(\vert 1,\downarrow\rangle - \vert -1,\downarrow\rangle)$ | $d$ |

II: Spin-orbit Contribution

Using the quantum state basis of $|L_z, S_z\rangle$ with $L_z \in \pm 2, \pm 1, 0$ in the d-orbitals, and $S_z \in \uparrow, \downarrow$ for spin-up and spin-down states, respectively, the effect of spin-orbit coupling term $\xi \vec{L} \cdot \vec{S} = \xi \left( L_z S_z + \frac{1}{2}(L_+ S_- + L_- S_+) \right)$ in terms of raising and lowering operators gives the interaction matrix in the z-basis:

$$H_{so,z} = \xi \begin{bmatrix} 1 & 0 & 0 & 0 & 0 & 0 & 0 & 0 & 0 & 0 \\ 0 & \frac{1}{2} & 0 & 0 & 0 & 1 & 0 & 0 & 0 & 0 \\ 0 & 0 & 0 & 0 & 0 & 0 & \frac{\sqrt{6}}{2} & 0 & 0 & 0 \\ 0 & 0 & 0 & -\frac{1}{2} & 0 & 0 & 0 & \frac{\sqrt{6}}{2} & 0 & 0 \\ 0 & 0 & 0 & 0 & -1 & 0 & 0 & 0 & 1 & 0 \\ 0 & 1 & 0 & 0 & 0 & -1 & 0 & 0 & 0 & 0 \\ 0 & 0 & \frac{\sqrt{6}}{2} & 0 & 0 & 0 & -\frac{1}{2} & 0 & 0 & 0 \\ 0 & 0 & 0 & \frac{\sqrt{6}}{2} & 0 & 0 & 0 & 0 & 0 & 0 \\ 0 & 0 & 0 & 0 & 1 & 0 & 0 & 0 & \frac{1}{2} & 0 \\ 0 & 0 & 0 & 0 & 0 & 0 & 0 & 0 & 0 & 1 \end{bmatrix}$$

Spin-orbit interaction matrix in the x-basis is formed by normal quantum mechanical coordinate transformation in terms of the already determined z-basis matrix. Explicitly:

$$H_{so,x} = \xi \begin{bmatrix} 0 & 0 & 0 & \frac{\sqrt{3}}{2} & 0 & 0 & 0 & 0 & 0 & -\frac{\sqrt{3}}{2} \\ 0 & 0 & 0 & 0 & \frac{1}{2} & 0 & 0 & 1 & \frac{1}{2} & 0 \\ 0 & 0 & 0 & \frac{1}{2} & 0 & 0 & 1 & 0 & 0 & \frac{1}{2} \\ \frac{\sqrt{3}}{2} & 0 & \frac{1}{2} & 0 & 0 & 0 & -\frac{1}{2} & 0 & 0 & \frac{1}{2} \\ 0 & \frac{1}{2} & 0 & 0 & 0 & \frac{\sqrt{3}}{2} & 0 & -\frac{1}{2} & \frac{1}{2} & 0 \\ 0 & 0 & 0 & 0 & \frac{\sqrt{3}}{2} & 0 & 0 & 0 & -\frac{\sqrt{3}}{2} & 0 \\ 0 & 0 & 1 & -\frac{1}{2} & 0 & 0 & 0 & 0 & 0 & -\frac{1}{2} \\ 0 & 1 & 0 & 0 & -\frac{1}{2} & 0 & 0 & 0 & -\frac{1}{2} & 0 \\ 0 & \frac{1}{2} & 0 & 0 & \frac{1}{2} & -\frac{\sqrt{3}}{2} & 0 & -\frac{1}{2} & 0 & 0 \\ -\frac{\sqrt{3}}{2} & 0 & \frac{1}{2} & \frac{1}{2} & 0 & 0 & -\frac{1}{2} & 0 & 0 & 0 \end{bmatrix}$$

Exact solutions of the full Hamiltonian are cumbersome, and we resort to the aforementioned numerical methods in general. Since the scale of spin-orbit coupling energy is small, a reasonable approximation can be obtained by treating SO coupling as a perturbation to the zeroth-order solutions in Table S6. Standard uses of degenerate and non-degenerate perturbation theory are necessary to approximate the perturbed states and energies. A summary of approximate states and energies along z, x directions are listed below.

**Table S6: summary of perturbed energies of real eigenstates. Difference between $E_z$, $E_x$ energies leads to preferential spin-alignment, and thus the anisotropy of magnetization.**

| State | Real State | Unperturbed Energy | $E_z$ Perturbation | $E_x$ Perturbation |
|---|---|---|---|---|
| $\boldsymbol{\varphi_1}$ | $\lvert \boldsymbol{z^2}, \uparrow\rangle$ | $A+D+2d$ | $\frac{3}{2}\xi^2\frac{1}{A+D+d}$ | $\frac{3}{4}\xi^2\left(\frac{1}{D+d}+\frac{1}{A+D+d}\right)$ |
| $\boldsymbol{\varphi_2}$ | $\lvert \boldsymbol{xy}, \uparrow\rangle$ | $A-2d$ | $\xi^2\left(\frac{1}{-D}+\frac{\frac{1}{2}}{A-3d}\right)$ | $\frac{3}{2}d+\frac{1}{2}\sqrt{9d^2+\xi^2}$ |
| $\boldsymbol{\varphi_3}$ | $\lvert \boldsymbol{x^2-y^2}, \uparrow\rangle$ | $A+D-2d$ | $\xi^2\left(\frac{1}{D}+\frac{\frac{1}{2}}{A+D-3d}\right)$ | $\xi^2\left(\frac{\frac{1}{4}}{A+D-3d}+\frac{\frac{1}{4}}{D-3d}+\frac{1}{A}\right)$ |
| $\boldsymbol{\varphi_4}$ | $\lvert \boldsymbol{xy}, \uparrow\rangle$ | $A+d$ | $\frac{1}{2}\xi$ | $\frac{1}{4}\xi\left(\frac{1}{A}+\frac{1}{A+3d}-\frac{1}{D-3d}-\frac{3}{D+d}\right)$ |
| $\boldsymbol{\varphi_5}$ | $\lvert \boldsymbol{yz}, \uparrow\rangle$ | $A+d$ | $-\frac{1}{2}\xi$ | $\frac{3}{2}d-\frac{1}{2}\sqrt{9d^2+\xi^2}$ |
| $\boldsymbol{\varphi_6}$ | $\lvert \boldsymbol{z^2}, \downarrow\rangle$ | $D+2d$ | $\frac{3}{2}\xi^2\frac{\frac{3}{2}}{-(A-D-d)}$ | $\frac{3}{4}\xi^2\left(\frac{1}{D-d}-\frac{1}{A-D-d}\right)$ |
| $\boldsymbol{\varphi_7}$ | $\lvert \boldsymbol{xy}, \downarrow\rangle$ | $-2d$ | $\xi^2\left(\frac{1}{-D}-\frac{\frac{1}{2}}{A+3d}\right)$ | $\frac{3}{2}d-\frac{1}{2}\sqrt{9d^2+\xi^2}$ |
| $\boldsymbol{\varphi_8}$ | $\lvert \boldsymbol{x^2-y^2}, \downarrow\rangle$ | $D-2d$ | $\xi^2\left(\frac{1}{D}-\frac{\frac{1}{2}}{A-D+3d}\right)$ | $\xi^2\left(\frac{\frac{1}{4}}{D-3d}-\frac{\frac{1}{4}}{A-D+3d}-\frac{1}{A-D}\right)$ |
| $\boldsymbol{\varphi_9}$ | $\lvert \boldsymbol{xy}, \downarrow\rangle$ | $d$ | $-\frac{1}{2}\xi$ | $d-\frac{1}{4}\xi^2\left(\frac{1}{A-3d}+\frac{1}{A}+\frac{3}{D-d}+\frac{1}{D-3d}\right)$ |
| $\boldsymbol{\varphi_{10}}$ | $\lvert \boldsymbol{yz}, \downarrow\rangle$ | $d$ | $\frac{1}{2}\xi$ | $\frac{3}{2}d+\frac{1}{2}\sqrt{9d^2+\xi^2}$ |

III: Numerical Method

In lieu of using perturbation theory, we used Python and pylab's computational package to diagonalize the Hamiltonian more generally. From this, the effect of strain is able to be more systematically tested based on relative displacements of the oxygen ligands in the $T_d$ and $O_h$ environments. For example, in octahedra the oxygen ligands are located at 6 positions relative to the center of the octahedron: $(\pm1,0,0), (0,\pm1,0), (0,0,\pm1)$. Strain is modeled through small displacement δ such that strain is applied in the appropriate plane, e.g. strained (001) NCO films experience displacements to $(1+\delta,0,0)$, $(-1-\delta,0,0)$, $(0,1+\delta,0)$, $(0,-1-\delta,0)$, $(0,0,1+2\delta)$, and $(0,0,-1-2\delta)$, under the assumption that the octahedra remain volume-invariant under the strain.

Direction of the electron spins is affected using the exchange interaction term of the Hamiltonian, $g\hbar\left(\vec{S_i}\cdot\vec{B}\right) = g\hbar\left(S_xB_x + S_yB_y + S_zB_z\right)$, with $S_i$ terms involving the Pauli spin matrices. Energies of the in-plane and out-of-plane spin directions are determined by diagonalizing the corresponding Hamiltonian, with the difference in energies being the anisotropy energy.

Matrix elements come directly from the crystal field matrix elements above (Equation S1), without regard for making symmetry arguments to determine which to measure. The effect of strain is linear in the difference between the c-axis expansion and the a-axis contraction like $d \sim \frac{U_{2,3,2}}{7}\frac{6(a-c)}{R}$. Numerical calculations to perform integrals remove the need for symmetry analysis, making the job much easier. Then, energies of the recombined states are calculated, and the cases between in-plane and out-of-plane energies as a function of $\delta$ can be compared, as shown in Fig. 4(c).

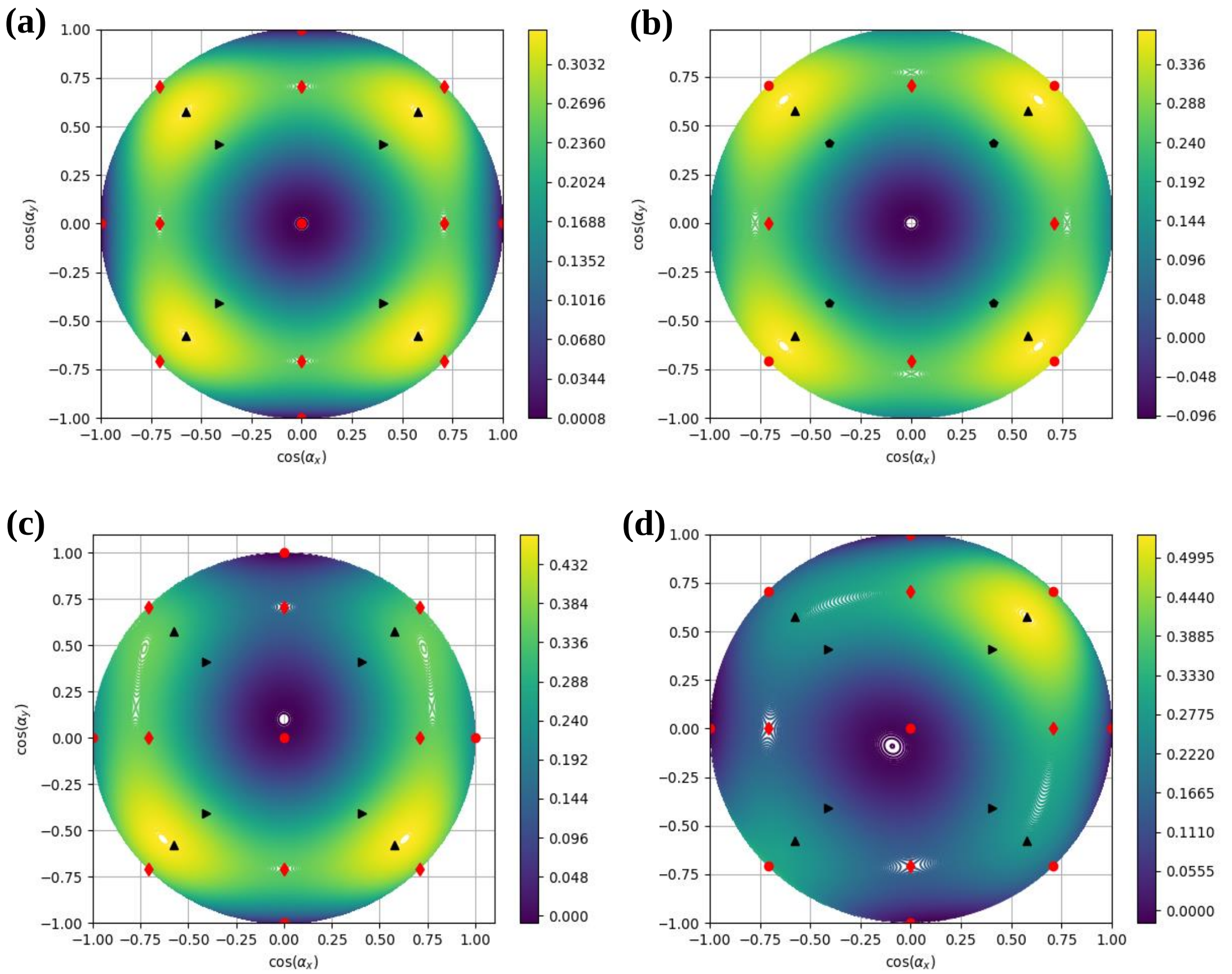

**Figure S4 | Single ion anisotropy simulation results.** Energy landscape solutions to the single-ion Hamiltonian, plotted as function of magnetization directions $\alpha_x$, $\alpha_y$. **(a)** is the unstrained Hamiltonian solution; **(b)-(d)** are the 0.3% compressively strained landscapes for strains in the (001), (110), and (111) crystal planes, respectively.